\documentclass[12pt]{article}

\usepackage[latin1]{inputenc}
\usepackage{epsfig}
\usepackage{a4wide}

\usepackage{latexsym}
\usepackage{amssymb}
\usepackage{amsmath}

\usepackage{graphics}

\begin{document}

\def\beq{\begin{equation}}
\def\eeq{\end{equation}}
\def\O{{\cal O}}
\def\N{{\cal N}}
\def\enu{\epsilon_\nu}
\def\hx{\hat\Delta}
\def\hatx{{\bf \hat x}}
\def\d{{\rm d}}
\def\i{{\rm i}}
\def\e{{\bf e}}
\def\ex{{\rm e}}
\def\x{{\bf x}}
\def\X{{\bf X}}
\def\J{{\rm J}}
\def\r{{\bf r}}
\def\s{{\bf s}}
\def\k{{\bf k}}
\def\y{{\bf y}}
\def\p{{\bf p}}
\def\q{{\bf q}}
\def\z{{\bf z}}
\def\R{{\bf R}}
\def\A{{\bf A}}
\def\v{{\bf v}}
\def\u{{\bf u}}
\def\w{{\bf w}}
\def\U{{\bf U}}
\def\V{{\bf V}}
\def\cm{{\rm cm}}
\def\l{{\bf l}}
\def\sec{{\rm sec}}
\def\Ckol{C_{Kol}}
\def\flux{\bar\epsilon}
\def\b{b_{kpq}}
\def\bfDelta{{\boldsymbol{\Delta}}}
\def\smali{{\scriptscriptstyle i}}
\def\smalfi{{\scriptscriptstyle \frac{5}{3} }}
\def\smalL{{\scriptscriptstyle{\rm L}}}
\def\smalP{{\scriptscriptstyle {\rm P}}}
\def\smalT{{\scriptscriptstyle {\rm T}}}
\def\smalE{{\scriptscriptstyle{\rm E}}}
\def\smal1n{{\scriptscriptstyle (1.1,n)}}
\def\smaln{{\scriptscriptstyle (n)}}
\def\smalA{{\scriptscriptstyle {\rm A}}}
\def\smalze{{\scriptscriptstyle (0)}}
\def\smalun{{\scriptscriptstyle 1}}
\def\smaldu{{\scriptscriptstyle 2}}
\def\smaln{{\scriptscriptstyle (n)}}
\def\smalel{{\scriptscriptstyle l}}
\def\gammaP{\gamma^\smalP}
\def\shell{{\tt S}}
\def\ball{{\tt B}}
\def\nav{\bar N}
\def\micron{\mu{\rm m}}
\font\brm=cmr10 at 24truept
\font\bfm=cmbx10 at 15truept

\baselineskip 0.7cm
\centerline{\brm Some relations between Lagrangian models}
\centerline{\brm and synthetic random velocity fields}
\vskip 20pt
\centerline{Piero Olla and Paolo Paradisi}
\vskip 5pt
\centerline{ISAC-CNR, Sezione di Lecce}
\centerline{Str. Prov. Lecce-Monteroni Km 1.2 I-73100 Lecce, Italy}
\vskip 20pt
\centerline{\bf Abstract}
\vskip 5pt
We propose an alternative interpretation of Markovian transport models based on the well-mixed
condition, in terms of the properties of a random velocity field with second order structure
functions scaling linearly in the space time increments. This interpretation allows direct 
association of the drift and noise terms entering the model, with the geometry of the 
turbulent fluctuations. In particular, the well known non-uniqueness problem in the
well-mixed approach is solved in terms of the antisymmetric part of the velocity
correlations; its relation with the presence of non-zero mean helicity and other 
geometrical properties of the flow is elucidated. The well-mixed condition appears
to be a special case of the relation between conditional velocity increments of the 
random field and the one-point Eulerian velocity distribution, allowing
generalization of the approach to the transport of non-tracer quantities. 
Application to solid particle transport leads to a model satisfying, in the 
homogeneous isotropic turbulence case, all the conditions on the behaviour 
of the correlation times for the fluid velocity sampled by the particles. 
In particular, correlation times in the gravity and in the inertia dominated case,
respectively, longer and shorter than in the passive tracer case; 
in the gravity dominated case, correlation 
times longer for velocity components along gravity, than for the perpendicular ones.
The model produces, in channel flow geometry, particle deposition 
rates in agreement with experiments.

\vfill\eject
\section*{I. Introduction}
\vskip 5pt
In Lagrangian models, the concentration of a quantity transported by a turbulent field is
reconstructed from the trajectories of the individual particles advected by the flow
\cite{taylor21,obukhov59,vandop85,thomson87}. 
Since each trajectory is associated with an independent 
realization of the turbulent flow, what is obtained is actually the mean concentration 
profile, given a certain distribution of sinks and sources for the transported quantity.

One keeps into account the fact that the turbulent length and time scales are non-zero,
by assuming that the 
particle velocity obeys a Langevin equation; this results in the system of equations, in 
terms of the particle Lagrangian velocity $\v$ and coordinate $\x$:
\beq
\begin{cases}
\d\x=\v\d t
\\
d\v ={\bf a}(\v,\x,t)\d t+\d{\bf w}
\\
\langle\d w^i\d w^j\rangle={\cal B}^{ij}(\v,\x,t)\d t
\end{cases}
\label{1.1}
\eeq
A strong physical motivation for a Lagrangian model in the form of Eqn.
(\ref{1.1})
is the linear scaling of the Lagrangian velocity structure functions for inertial range time 
separations \cite{pope}. In the case of passive tracers, we have
\beq
\langle [v^i(t)-v^i(0)][v^j(t)-v^j(0)]\rangle\simeq \delta^{ij}C_0\bar\epsilon t
\label{1.2}
\eeq
with $\bar\epsilon$ the mean viscous dissipation and $C_0$ a universal constant;
Eqn. (\ref{1.1}) will then result from assuming a Markovian behaviour for $\v$, i.e. 
that $a^i$ and ${\cal B}^{ij}$ depend solely on the current values of $\v$ and $\x$ and not 
on their previous history. 

The well-mixed condition, introduced in \cite{thomson87}, lead to a great
advance in Lagrangian models, providing a simple technique for expressing the
drift coefficient ${\bf a}$ in Eqn. (\ref{1.1}) in terms of observed properties
of the flow. 

In the current approach, it is assumed
that the noise term $\d\w$ in Eqn. (\ref{1.1}) accurately represents, in high Reynolds number 
turbulent regimes, the inertial range scaling of the Lagrangian velocity increments:
\beq
{\cal B}^{ij}(\v,\x,t)=\delta^{ij}C_0\bar\epsilon(\x,t)
\label{1.3}
\eeq
Then, at least when the flow is incompressible, 
the well-mixed criterion allows us to determine the drift coefficient
${\bf a}$ in terms of $\overline{\epsilon}$ and the Eulerian Probability Density Function 
(PDF) for the fluid velocity $\u(\x,t)$: $\rho_E(\u,t,\x)\equiv\rho(\u(\x,t))$. 
(In the general compressible case, the Eulerian PDF must be weighed on the
fluctuating fluid density, which is tantamount to substitute $\rho_E$ with the 
Lagrangian PDF for the fluid parcels velocity and position \cite{thomson87}).

The great advantage of the well-mixed approach, coupled with Eqn.
(\ref{1.3}), is that no 
knowledge of the spatio-temporal structure of the turbulent fluctuations is required, rather, 
it is the outcome, encoded in the drift coefficient, of the well mixed condition. 
This strength of the model, however,
turns into weakness, whenever the turbulent structure plays a relevant role. 
A first hint that this, actually, is always the case, is the non-uniqueness of the
solution for ${\bf a}$ given $\bar\epsilon$ and $\rho_E(\u)$ \cite{thomson87,sawford99}. 
In the well mixed approach, the 
drift coefficient ${\bf a}$ is determined only up to a velocity curl, and interpretation of this
freedom in terms of the properties of the flow is awkward. 

There are situations in which the structure of turbulence plays an explicit role. A first
example is produced when coherent structures dominate the flow, a rather common occurrence
in turbulence, which takes a dramatic form in near wall regions.
These regions become relevant in many situations of practical interest in industrial flows, 
but also, to name a few, in the study of transport in tree canopies and in indoor pollution.
These flows are often characterized by moderate Reynolds number, and, for this reason, not only 
the viscous scale may be not negligible, but a well developed inertial range may even be absent,
so that the conditions justifying Eqn. (\ref{1.3}) cease to be valid.
Now, standard techniques exist which allow for the inclusion of
non-Gaussianity \cite{luhar89} and anisotropy \cite{borgas97,sawford99} in Lagrangian models,
as well as for the effect of finite Reynolds numbers \cite{sawford91a}.
However, these techniques do not take into account the 
geometric structure of the turbulent fluctuations, which calls for information about space 
correlation.

In the range of scales we are considering, another issue will come into play, if we are
interested in modelling solid particle transport. In this case, non-tracer
behaviours associated with inertia and gravity will begin to be felt 
(inertia and trajectory crossing effects 
\cite{csanady63,reeks77,sawford91}).
This is especially true for atmospheric aerosol, characterized by heavy 
particles with relative density  of the order of $1000$ and particle
diameters in the range $10^{-2} \div 10^2 \mu {\rm m}$. 
Inertia effects are generally negligible in
ABL (atmospheric boundary layer) mesoscale modeling, 
but can become relevant in the near wall regions of
wall bounded flows \cite{friedlander57},
which is relevant for problems of
air conditioning and abatement of indoor pollution.
Trajectory crossing effects related to gravity can have deep implications not
only in wall bounded flows, but also for particulate transport in the ABL
\cite{csanady63}.

These effects reflect heavily on the
possibility of using Eqns. (\ref{1.2}-\ref{1.3}) to model the Lagrangian
velocity increments, and call back for the need of information on the
spatio-temporal structure of the turbulent fluctuations.

Some attempts in this direction were carried on in \cite{desjonqueres88,berlemont90}, 
but disregard of correlations between fluid velocity increments along solid particle 
trajectories lead to difficulties in the fluid particle limit \cite{olla02} 
and in the implementation of the well-mixed condition.

\vskip 5pt
In order to be able to understand the constraints imposed by the turbulent structure 
on the form of a Lagrangian model, one may try a derivation from
a velocity field $\u(\x,t)$ of prescribed statistics. 
If the structure function $\langle [u^i(\x,t)-u^i(0,0)][u^j(\x,t)-u^j(0,0)]\rangle$
scaled linearly for small space-time separations, the velocity increment
between two points lying on a trajectory would be given by:
\beq
\d\u=\langle[\partial_t+\u(\x,t)\cdot\nabla]\u(\x,t)|\u(\x,t)\rangle\d t+\d\w
\label{1.4}
\eeq
with $\langle\d\w\rangle=0$ and $\d w^2=O(\d t)$. We could then introduce a Lagrangian model
obeying Eqn. (\ref{1.1}), with
\beq
\begin{cases}
{\bf a}(\v,\x,t)=\langle[\partial_t+\u(\x,t)\cdot\nabla]\u(\x,t)|\u(\x,t)=\v\rangle
\\
{\cal B}^{ij}(\v,\x,t)\d t=\langle\d w^i\d w^j|\u(\x,t)=\v\rangle
\end{cases}
\label{1.5}
\eeq
With the coefficients given in this equation, Eqn. (\ref{1.1}) would provide a Markovianized 
version of the dynamics of a particle moving in the random velocity field $\u(\x,t)$. 

It turns out that, provided the structure functions for $\u$ scale linearly, the well-mixed 
technique could be imposed directly on the random field $\u(\x,t)$, before any trajectory is
defined. This means expressing the form of the conditional averages of the velocity derivatives:
$\langle\partial_t\u(\x,t)|\u(\x,t)\rangle$ and $\langle\nabla\u(\x,t)|\u(\x,t)\rangle$ 
in terms of the Eulerian PDF $\rho_E(\u,\x,t)$. This has the important consequence
that a passive tracer advected by an incompressible flow, will satisfy
the ergodic property
\beq
\rho_L(\v|\x,t)=\rho_E(\v,\x,t)
\label{1.6}
\eeq
where $\rho_L(\v|\x,t)$ is the Lagrangian PDF for a tracer passing at the point
$\x$ at time $t$, to have 
velocity $\v$. Thus, the well-mixed condition imposed 
on the random field extends naturally to the Lagrangian model defined by Eqns. 
(\ref{1.1}) and (\ref{1.5}). This is advantageous in the compressible case, where it 
will be shown that, contrary to the Thomson-87 approach, knowledge of $\rho_E$ is 
sufficient for the determination of a well-mixed model.

\vskip 5pt
Clearly, linear scaling at small separations does not correspond to the properties of a real 
turbulent field, which, at high Reynolds numbers is more rough, and whose time correlations, 
due to the sweep effect, have Lagrangian nature at short time scales \cite{pope}.
This is compensated, however, by control over the large scale 
structure of the correlations, which is the relevant aspect for the determination of turbulent 
transport. 

A related issue, concerning solid particle transport, is that anomalous scaling of the fluid 
velocity increments sampled by a solid particle,
are known to occur at sufficiently short time scales \cite{shao95}. 
Analysis of the different ranges characterizing solid particle motion was carried on in 
\cite{olla02}, and Lagrangian models resolving the anomalous scaling range were presented
in \cite{reynolds02} and \cite{olla02}, based respectively on the use of fractional 
Brownian motion and synthetic turbulence algorithms. Again, consideration of these
short-time effects is neglected in favour of control of large scale geometry.

Compared to the standard approach in Lagrangian modelling, the one proposed
here has definite advantages.
Spatio-temporal turbulent structures can be included in a relatively simple
way.
The non-uniqueness problem is solved in a simpler way, 
since only purely Eulerian properties of the flow are invoked
(helicity is one example). 
The advection of passive tracers and solid particles are treated exactly 
on the same footing, hence, extension of the model to solid particle transport is automatic and 
does not need introducing additional assumptions.

\vskip 5pt
This paper is organized as follows. 
In section II, a local characterization of a random velocity
field will be given, introducing generalized ''four-dimensional'' Langevin and Fokker-Planck
equations, and providing local and global existence conditions.
The condition of local existence appears to take the form of
a generalized form of the well-mixed condition, which will be discussed in 
section III. This will be used
to calculate conditional averages in the form $\langle\nabla\u|\u\rangle$ and 
$\langle\partial_t\u|\u\rangle$ from the property of $\d\w$ and the Eulerian velocity
PDF, and the relation with the spatio-temporal structure of the random field will be discussed.
In section IV, expressions for the noise amplitude $\langle\d\w\d\w\rangle$ will be
derived, and their relation with the symmetric sector of the velocity correlation will
be discussed in terms of the SO(3) technique introduced in \cite{arad99}.
The antisymmetric sector of the velocity correlation will be discussed in section V, 
illustrating how it relates to the problem of non-uniqueness in the well-mixed
approach, and showing how helicity and other geometrical features could be included 
in the random field.
Section VI will be devoted to the derivation of a Markovian Lagrangian model in the form 
of Eqn. (\ref{1.5}), and to presentation of its main properties. The relation with 
the Thomson-87 model \cite{thomson87} will be discussed.
Sections VII and VIII will be devoted to analysis of the Markovian approximation in the
Lagrangian model and to proof of the ergodic property given by Eqn. (\ref{1.6}).
Sections IX and X will illustrate two applications of the Lagrangian model to solid
particle transport, respectively, in homogeneous isotropic turbulence, and in a 
turbulent channel flow.
Section XI contains the conclusions.

\vskip 20pt
\section*{ II. Characterization of the random velocity field}
\vskip 10pt
Let us introduce a zero-mean, incompressible random velocity field $\u(\x,t)$, with 
2nd order structure functions scaling linearly
in the increment at small space-time separations. 
We introduce 4-vector notation:
\beq
x^\mu=\{x^0,x^i\}\equiv\{t,\x\}, 
\qquad
\partial_\mu\equiv\frac{\partial}{\partial x^\mu}.
\label{2.1}
\eeq
and stick rigorously to the Einstein convention of summation over covariant-contro- variant 
repeated indices.
We have the following equation for the velocity increment:
\beq
\d u^i\equiv\d x^\mu\partial_\mu u^i
={A_\mu}^i(\u,x^\mu)\d x^\mu+\d w^i
\label{2.2}
\eeq
where 
\beq
{A_\mu}^i(\u,x^\mu)=\langle\partial_\mu u^i(\x,t)|\u(\x,t)\rangle
\label{2.3}
\eeq
and $\langle\d\w|\u\rangle=0$. From linearity, the contribution to the velocity structure 
function is dominated, for small values of the increments, by the correlation for $\d\w$, 
and we have:
\beq
\langle\d w^i\d w^j|\u(\x,t)\rangle=\langle\d u^i(\x,t)\d u^j(\x,t)\rangle=O(|\d\x|,\d t)
\label{2.4}
\eeq
We limit our analysis to random velocity fields
where the statistics of the velocity increments is independent of that of the total velocity:
\beq
\langle\d w^i\d w^j|\u\rangle=\langle\d w^i\d w^j\rangle.
\label{2.5}
\eeq
Incompressibility of the velocity field $\partial_iu^i=0$, leads to the constraint,
indicating $\Delta^\mu\equiv\Delta x^\mu$:
\beq
A^i_i=0,
\qquad
\frac{\partial\langle\Delta w^i\Delta w^j\rangle}{\partial\Delta^i}=0
\label{2.6}
\eeq
From Eqns. (\ref{2.2}) and (\ref{2.4}), we obtain the following generalized It\^o's Lemma; 
for a generic smooth function $\phi(\u)$:
\beq
\d\phi(\u)=(A_\mu^{\ i}\d x^\mu+\d w^i)\partial_{u^i}\phi(\u)+\frac{1}{2}\langle\d w^i\d w^j\rangle
\partial_{u^i}\partial_{u^j}\phi(\u)
\label{2.7}
\eeq
and from here, we can derive an equation for the change of the 1-point PDF 
$\rho_E(\u,x^\mu)\equiv\rho(\u(\x,t))$, in passing from the point $x^\mu$ to to the point 
$x^\mu+\d x^\mu$:
\beq
\d\rho_E\equiv\d x^\mu\partial_\mu\rho_E=-\partial_{u^i}(A_\mu^{\ i}\d x^\mu\rho_E)
+\frac{1}{2}\partial_{u^i}\partial_{u^j}(\langle\d w^i\d w^j\rangle\rho_E)
\label{2.8}
\eeq
Notice that the form of the two equations (\ref{2.7}) and (\ref{2.8}) is independent of 
incompressibility and Eqn. (\ref{2.6}).
The sequence leading from Eqn. (\ref{2.2}) to (\ref{2.8}) is very suggestive, in that it 
generalizes the one from a Langevin to a Fokker-Planck equation \cite{gardiner}. However, 
contrary to the 
case of a standard 
Fokker-Planck equation, Eqn. (\ref{2.8}) does not admit in general solution for $\rho_E$.
In fact, once the noise amplitude
$\langle\d w^i\d w^j\rangle$ and the drift ${A_\mu}^i$ are given, Eqn. 
(\ref{2.8}) becomes
a system of four partial differential equations for the single PDF $\rho_E$, and 
this system is generally over-determined. In the next section, it will be shown how a 
generalized version of the well-mixed condition is able to take care of this local
existence problem.

The ill-posedness of the problem is reflected at the global level, in the fact that
a local definition for the ''noise'' increment amplitude $\langle\d w^i\d w^j\rangle$ is
not sufficient to define a realization for $\w(\x,t)$, and consequently for $\u(\x,t)$. 
This in contrast with the case of the standard Langevin equation. In fact,
if we integrate Eqn. (\ref{2.2}) along a closed curve in space-time, and consider uncorrelated 
increments $\d\w$ along the curve, we will obtain in general a non-zero total velocity increment
in the closed loop.
In other words, if we disregard these correlations for $\w$, the differential $\d\u$ entering
Eqn. (\ref{2.2}) will not in general be exact.

The question becomes at this point the existence of a random velocity field with local structure
described by Eqn. (\ref{2.2}). It turns out that such a velocity 
field can be constructed explicitly, although the construction described below is by no means 
unique.

Given a point $x^\mu$ and a direction in space time defined by the versor $r^\mu$ 
($r^\mu r_\mu=1$),
we can introduce the stochastic process $\hat u^i(s)\equiv\hat u^i(x^\mu,r^\mu;s)$ obeying the 
Langevin equation:
\beq
\begin{cases}
\d \hat u^i(s)=r^\mu{A_\mu}^i(\hat\u,x^\mu)\d s+\d\hat w^i
\\
\langle\d\hat w^i\d\hat w^j)\rangle=\frac{\d}{\d s}
\langle[u^i(x^\mu+r^\mu s)-u^i(x^\mu)][u^j(x^\mu+r^\mu s)-u^j(x^\mu)]
\rangle\Big|_{s=0^+}\d s
\end{cases}
\label{2.9}
\eeq
The correlation functions for the stochastic process $\hat\u(s)$, starting from the second
order one 
$C^{ij}(x^\mu,sr^\mu)=\langle\hat u^i(x^\mu,r^\mu;-s/2)\hat u^j(x^\mu,r^\mu;s/2)\rangle$, 
will identify a  
random velocity field $\u(x^\mu)$ whose local statistical properties are those imposed by
Eqn. (\ref{2.2}), and whose restriction to straight lines in space time will be, by 
construction, Markovian. As with the correlation time of the solution of a standard 
Langevin equation, the correlation length in the direction $r^\mu$ will be encoded 
in the drift coefficient $r^\mu{A_\mu}^i(\hat\u,x^\mu)$.
A random field realization is obtained, in the simpler Gaussian case, by first 
carrying on the principal orthogonal decomposition (POD) of $C^{ij}(x^\mu,\Delta^\mu)$, 
and then random superposing, with the appropriate amplitudes, the resulting POD modes
\cite{holmes}.

\vskip 20pt
\section*{ III. Determination of the drift}
\vskip 5pt
The real meaning of Eqn. (\ref{2.8}) is to provide a consistency condition for 
$\langle\d w^i\d w^j\rangle$ and ${A_\mu}^i$, that could be used to generalize the
Thomson-87 technique and determine from $\langle\d w^i\d w^j\rangle$ and $\rho_E$, the 
expression for ${A_\mu}^i$. The difference with the standard case is that, instead 
of calculating the conditional mean $\langle\partial_tv^i|\v\rangle$ of the Lagrangian velocity
time derivative, we seek here the conditional mean of all the derivatives of the Eulerian
velocity, namely $\langle\partial_\mu u^i|\u\rangle$. These averages contain important information
on the behaviour of the velocity correlation
$C^{ij}(x^\mu,\Delta^\mu)=\langle u^i(x^\mu-\Delta^\mu/2)u^j(x^\mu+\Delta^\mu/2)\rangle$:
\beq
C^{ij}(x^\mu,\Delta^\mu)=\frac{1}{2}[R^{ij}(x^\mu-\Delta^\mu/2)+R^{ij}(x^\mu+\Delta^\mu/2)]
-\frac{1}{2}\langle\Delta u^i\Delta u^j\rangle
+C^{ij}_A
\label{3.1}
\eeq
Here, $R^{ij}(x^\mu)=C^{ij}(x^\mu,0)$ indicates the Reynolds tensor, while
\beq
C_A^{ij}=\frac{1}{2}[C^{ij}(x^\mu,\Delta^\mu)-C^{ij}(x^\mu,-\Delta^\mu)]
\label{3.2}
\eeq
is the antisymmetric part of the velocity correlation. It is clear that the noise amplitude is 
associated with the symmetric sector of the velocity correlation, and for small $\Delta^\mu$: 
$\langle\Delta w^i\Delta w^j\rangle\simeq\langle\Delta u^i\Delta u^j\rangle$. 

Let us try to generalize the
Thomson-87 approach to calculate the drift ${A_\mu}^i$ from $\rho_E$.
It is convenient to split the drift into three pieces:
\beq
A_\mu^{\ i}=\bar A_\mu^{\ i}+\frac{1}{\rho_E}\Phi_\mu^{\ i}+\frac{1}{\rho_E}\Psi_\mu^{\ i}
\label{3.3}
\eeq
where $\bar A_\mu^{\ i}$ is chosen to cancel
the noise term in the Fokker-Planck equation (\ref{2.8}).
Exploiting independence of the noise amplitude from $\u$:
\beq
\bar A_\mu^{\ i}\d x^\mu=\frac{1}{2}\langle\d w^i\d w^j\rangle\partial_{u^j}\log\rho_E
\label{3.4}
\eeq
and $\bar A_\mu^{\ i}$, from the second of Eqn. (\ref{2.6}), is automatically traceless.
The term ${\Phi_\mu}^i$ is chosen to cancel the contributions to Eqn.
(\ref{2.8})
from statistical non-uniformity and non-stationarity:
\beq
\partial_{u^i}\Phi_\mu^{\ i}=-\partial_\mu\rho_E
\label{3.5}
\eeq
The term $\Psi$ is necessary to cancel the trace of $\Phi$ and must be divergenceless with
respect to $\u$:
\beq
\partial_{u^i}\Psi_j^{\ i}=0,
\qquad
{\Psi_0}^i=0
\qquad
\Psi_i^i=-\Phi^i_i
\label{3.6}
\eeq
As in the Thomson-87 approach \cite{thomson87}, the drift is defined up to a non-unique term 
$\frac{1}{\rho_E}{\Xi_\mu}^i$
satisfying: 
\beq
\Xi^i_i=0,
\qquad
\partial_{u^i}{\Xi_\mu}^i=0
\label{3.7}
\eeq
which, substituted into Eqn. (\ref{2.8}), will
produce an identically zero contribution.

The drift $A_\mu^{\ i}$ is associated with the velocity correlation through the equation:
\beq
\langle u^i\Delta u^j\rangle=\langle u^i\langle\Delta u^j|\u\rangle\rangle
=\langle u^i{A_\mu}^j\rangle\Delta^\mu,
\label{3.8}
\eeq
Let us analyze individually each of the terms in $A_\mu^{\ i}$. Substituting Eqn. (\ref{3.4}) into
Eqn. (\ref{3.8}) we see that $\bar A_\mu^{\ i}$ gives just the symmetric piece of the correlation, 
i.e.  the $-\frac{1}{2}\langle\Delta u^i\Delta u^j\rangle$  in Eqn. (\ref{3.8}). 
The $\Phi$ and $\Psi$
terms are more easily analyzed in Fourier space: $\tilde f({\boldsymbol{\eta}})=
\int\d^3u\ex^{-\i\eta_iu^i}f(\u)$. Using Eqns. (\ref{3.3}-\ref{3.4}), Eqn. (\ref{3.8}) 
will read:
\beq
\langle u^i\Delta u^j\rangle=-\frac{1}{2}\langle\Delta w^i\Delta w^j\rangle
-\i\Delta^\mu\partial_{\eta_j}(\tilde\Phi_\mu^{\ i}+\tilde\Psi_\mu^{\ i})|_{\eta=0}
\label{3.9}
\eeq
Using the fact that the generating function $\tilde\rho_E$
obeys $\tilde\rho_E=1-\frac{1}{2}R^{ij}\eta_i\eta_j+O(\eta^3)$, we can write, from Eqn. 
(\ref{3.8}):
\beq
\tilde\Phi_\mu^{\ i}=-\frac{\i}{2}\partial_\mu R^{ij}\eta_j+O(\eta^2)
\label{3.10}
\eeq
so that the contribution from $\Phi$ to the correlation function is, from Eqn. (\ref{3.9}):
$\frac{1}{2}\d x^\mu\partial_\mu R^{ij}$, which accounts for the spatial inhomogeneity of
the correlation $[$the $\frac{1}{2}(R^{ij}+R^{ij})$ term on RHS of Eqn.
(\ref{3.1}), which is centered at $\sim\ x^\mu]$.

We see that $\bar A_\mu^{\ i}$ and $\Phi_\mu^{\ i}$ account for all of the contribution to
the correlations, which either are symmetric, or come from inhomogeneity of the
statistics. We know at this point that both $\Psi_\mu^{\ i}$ and ${\Xi_\mu}^i$ will be
able to contribute only to the antisymmetric part of ${A_\mu}^i$. We give in explicit
form the contribution from $\Psi_\mu^{\ i}$. Exploiting the first of Eqn. (\ref{3.6}),
we utilize the ansatz:
\beq
\tilde\Psi_j^{\ i}=\i (\delta^i_j\eta_k\tilde\psi^k-\eta_j\tilde\psi^i),
\qquad
\tilde\Psi_0^{\ i}=0
\label{3.11}
\eeq
and, from $\Psi^i_i=-\Phi^i_i$ and Eqn. (\ref{3.10}):
\beq
\tilde\psi^k=\frac{1}{4}(\partial_lR^{lk})+\epsilon^{klm}\eta_l\tilde f_m+O(\eta^2)
\label{3.12}
\eeq
with ${\bf f}$ arbitrary. The contribution to ${\Psi_j}^i$ from ${\bf f}$ is traceless  
and can be reabsorbed into the non-unique term $\Xi$; the final result is therefore:
\beq
\tilde\Psi_j^{\ i}=\frac{\i}{4}[\delta^i_j(\partial_lR^{lk})\eta_k
-(\partial_lR^{li})\eta_j]+O(\eta^3)
\label{3.13} 
\eeq
and the contribution to the correlation function is:
$\frac{1}{4}[\partial_lR^{lk}\d x^i-\partial_lR^{li}\d x^k]$, which is antisymmetric 
as required.

Explicit expressions for the drift terms are promptly obtained in the case of Gaussian
statistics (expressions for the case of a symmetric $\rho_E$ with kurtosis 
larger than three are given in the Appendix A). The velocity PDF reads therefore:
\beq
\rho_E(\u,x^\mu)\equiv\rho_G(\u,x^\mu)=(8\pi^3||R||)^{-\frac{1}{2}}\exp(-\frac{1}{2}S_{ij}u^iu^j)
\label{3.14}
\eeq
with $S_{ij}=(R^{-1})_{ij}$ the Reynolds tensor inverse. 
In this case, the higher order terms in $\eta_i$ entering Eqns.
(\ref{3.10}-\ref{3.13}) disappear and we are left with:
\beq
\bar A_\mu^{\ i}\d x^\mu=-\frac{1}{2}\langle\d w^i\d w^j\rangle S_{jk}u^k
\label{3.15}
\eeq
\beq
\Phi_\mu^{\ i}=\frac{1}{2}(\partial_\mu R^{ik})S_{kl}u^l\rho_E
\label{3.16}
\eeq
\beq
\Psi_j^{\ i}
=\frac{1}{4}[-\delta^i_j(\partial_lR^{lk})S_{km}+(\partial_lR^{li})S_{jm}]u^m\rho_E
\label{3.17}
\eeq
We can use these explicit expressions to obtain more informations on the nature of the various 
contributions to the drift. 
In analogy to the case of the standard Langevin equation, we see that
$\bar A_\mu^{\ i}$ must be discontinuous at $\Delta^\mu=0$.
From Eqn. (\ref{3.4}), discontinuity of the correlation function derivative at $\Delta^\mu=0$ is 
necessary to balance the linear scaling of $\d w^2$. 
In the coordinate system where, for the given $\Delta^\mu$,
$\bar A_j^{\ i}$ is diagonal, we shall then have:
\beq
\langle\Delta u^i|\u\rangle\sim-|\Delta^i|
\label{3.18}
\eeq
As regards $\Phi_\mu^{\ i}$, we see from Eqn. (\ref{3.16}) that it produces an amplification of
$\u$ when $\d x^\mu$ is directed to a region in space-time where the turbulence is stronger
(this is easy to see when $R^{ij}\propto\delta^{ij}$).

Finally, the $\Psi$ term, turns out to produce a complicated mixture of rotations
and amplification of the velocity vector. 
Indicating $\tilde u_i=S_{ik}u^k$, and
choosing the coordinate system so that 
$\partial_iR^{il}=4\tilde c\delta^l_1$, and $\Delta^3=0$
we have from Eqn. (\ref{3.17}):
\beq
\langle\Delta u^1|\u\rangle=\tilde c\tilde u_2\Delta^2,
\quad
\langle\Delta u^2|\u\rangle=-\tilde c\tilde u_1\Delta^2,
\quad
\langle\Delta u^3|\u\rangle=0.
\label{3.19}
\eeq
If $R^{ij}\propto\delta^{ij}$, we will have $\tilde u_i=u_i$ and the result of Eqn. (\ref{3.19}) 
will be a rotation of $\u$ in the plane $12$ as one moves in the direction $x_2$.

\vskip 20pt
\section*{IV. Determination of the noise tensor}
\vskip 5pt
In order to obtain the drift coefficients, which give the decay of the turbulent correlations
in the various space-time directions, it is necessary first to determine the form of the
noise tensor $\langle\d w^i\d w^j\rangle$. In fact, it is in the noise that all the information
on the turbulent structure is encoded (at least that part relative to the symmetric sector of 
the correlations). In the case of a Gaussian random velocity field, the noise tensor can be
determined directly from the turbulent correlations by means of a fit in terms of products
of exponentials with sines and cosines (a common practice in turbulence theory; consider e.g.
the Frenkiel functions \cite{hinze}). Indicating $\d x^\mu=r^\mu\d s$, $r^\mu r_\mu=1$,
we fit the turbulent correlation by the expression:
\beq
\frac{\partial}{\partial s}\langle u^i(x^\mu)u^j(x^\mu+r^\mu s)\rangle=
{c_k}^j\langle u^i(x^\mu)u^k(x^\mu+r^\mu s)\rangle,
\label{4.0.1}
\eeq
where ${c_i}^j$ depends on the direction $r^\mu$, the mid-point position $x^\mu+\Delta^\mu/2$, 
but not on $\d s$.
This imposes linear dependence of the random field drift on the velocity:
\beq 
\langle\d u^i|\u\rangle=\d x^\mu{A_\mu}^i={c_j}^iu^j\d s
\label{4.0.0}
\eeq 
(notice that Gaussian statistics, by itself, imposes linearity
through the well mixed condition, only on the symmetric contribution to the drift 
${\bar A_\mu}^i$).
Using Eqn. (\ref{4.0.0}), Eqn. (\ref{3.8}) takes the form:
$\langle u^k\d u^i\rangle= {c_j}^iR^{jk}\d s$
and, from Eqn. (\ref{3.2}), we obtain:
\beq
\langle\d w^i\d w^j\rangle=\langle\d u^i\d u^j\rangle=
\frac{1}{2}({c_k}^iR^{kj}+{c_k}^jR^{ki})\, \d s
\label{4.0.3}
\eeq
We stress that, although Eqn. (\ref{4.0.3}) describes the behaviors of the random field
correlations at small separations, the coefficients ${c_j}^i$ descend from a fit
of turbulent correlations at finite separations.

A very general form for the noise tensor, satisfying the incompressibility condition
$\partial\langle\Delta w^i\Delta w^j\rangle/\partial\Delta^i=0$, 
allowing association of Eqn. (\ref{4.0.3}) with geometric features of the flow, 
is a superposition of terms in the form:
\beq
\langle\Delta w^i\Delta w^j\rangle=\frac{2u_T^2}{\tau_E}[B_t^{ij}(\Delta^0)
+B^{ij}(\bfDelta-\bar\u\Delta^0))];
\qquad
\partial_iB^{ij}(\bfDelta)=0.
\label{4.1}
\eeq
Here, 
$\tau_E$ fixes the time scales of the fluctuations
and in the Gaussian case coincides with the Eulerian correlation time (see next), 
$u_T^2=\frac{1}{3}R^i_i$, and
$B_t^{ij}=|\Delta^0|\delta^{ij}+\hat B_t^{ij}$, with $\hat B_t^{ij}$ symmetric and traceless. 
(For lighter notation we leave in this section the dependence from the space-time position
unindicated).

We see that the presence of mixed space-time increment contributions can account for 
situations in which the time correlations have Lagrangian nature. 
In this way, pure time decorrelation will take place
in the reference system moving locally with the mean flow $\bar\u$.
A situation with purely Eulerian time correlation will be realized by putting $\bar\u=0$.


In moderately anisotropic situations, it may be expedient to expand the space component 
$B^{ij}$ in spherical tensors, following the SO(3) decomposition technique \cite{arad99}: 
\beq
B^{ij}(\bfDelta)=\sum_{J=0}B^{ij}_J(\bfDelta)
\label{4.3}
\eeq
where $B^{ij}_J$ indicates a combination of $J$-th order spherical tensors (see Appendix B). 
The symmetry of $\langle\Delta w^i\Delta w^j\rangle$ imposes selection rules on which 
spherical tensors may contribute; it turns out that to keep 
the lowest order anisotropic contribution, it is enough to consider
spherical tensors of order $J=0$ and $J=2$.  
The incompressibility condition $\partial_{\Delta^i}B^{ij}_J=0$ gives then (the hats identify versors):
$$
B^{ij}(\bfDelta)=\frac{|\bfDelta|}{u_T}[(a+4b^{lm}\hx_l\hx_m)\delta^{ij}+\frac{1}{3}(-a+(2b^{lm}-c^{lm})
\hx_l\hx_m)\hx^i\hx^j
$$
\beq
-\hx_l[(2b^{li}+c^{li})\hx^j+(2b^{lj}+c^{lj})\hx^i]+4c^{ij}]
\label{4.4}
\eeq
where $a$ gives the $J=0$ part, while the tensors $b^{ij}$ and $c^{ij}$, which are symmetric
and traceless, account for the the $J=2$ part.
We consider next some relevant limit cases.

\vfill\eject
\noindent{\it Isotropic turbulence}
\vskip 5pt
In this case, all the spherical tensors with $J>0$
are zero. We are thus left with the simple expression:
\beq
\langle\Delta w^i\Delta w^j\rangle=\frac{2u_T^2}{\tau_E}[|\Delta^0|\delta^{ij}
+\frac{a|\bfDelta|}{u_T}(\delta^{ij}-\frac{1}{3}\hx^i\hx^j)]
\label{4.5}
\eeq
The parameter $a$ identifies a length-scale $l_u=u_T\tau_E/a$ for the fluctuations and has 
therefore the meaning of a ratio between the eddy life-time $\tau_E$ and the eddy rotation
time $l_u/u_T$.

\vskip 10pt
\noindent{\it Long axisymmetric vortices}
\vskip 5pt
Let us imagine that the correlation tensor is dominated by 
the effect of long axisymmetric
vortices directed along $x^1$. Let us try to use this
information to impose a structure to the space structure tensor $B^{ij}$ 
defined in Eqn. (\ref{4.4}).
Let us impose the condition that $B^{ij}(\bfDelta)=0$ for $\bfDelta=\{\Delta,0,0\}$. For 
$\bfDelta=\{\Delta,0,0\}$,
we have from Eqn. (\ref{4.4}):
$$
\frac{u_TB^{11}}{|\bfDelta|}=\frac{1}{3}(2a+2b^{11}+5c^{11}),
\quad
\frac{u_TB^{22}}{|\bfDelta|}=a+4b^{11}+4c^{22},
\quad
\frac{u_TB^{33}}{|\bfDelta|}=a+4b^{11}+4c^{33},
$$
\beq
\frac{u_TB^{12}}{|\bfDelta|}=3c^{12}-2b^{12},
\quad
\frac{u_TB^{13}}{|\bfDelta|}=3c^{13}-2b^{13},
\quad
\frac{u_TB^{23}}{|\bfDelta|}=4c^{23}.
\label{4.6}
\eeq
and we find immediately the result:
$$
c^{12}=c^{13}=\frac{2}{3}b^{12}=\frac{2}{3}b^{13};
\qquad
c^{23}=0;
$$
\beq
b^{11}=-\frac{3}{8}a;
\qquad
b^{22}=b^{33}=\frac{3}{16}a;
\qquad
c^{11}=-\frac{a}{4};
\qquad
c^{22}=c^{33}=\frac{a}{8}.
\label{4.7}
\eeq
We are free to impose the condition $b^{1i}=c^{1i}=0$ for $i\ne 1$ and 
we reach the expression for generic $\bfDelta$ of the $B$ 
components along $11$ and $22$:
\beq
\begin{cases}
\frac{u_TB^{11}(\bfDelta)}{|\bfDelta|}=\frac{1}{6}\hx_1^2+\frac{3}{4}\hx_\perp^2
-\frac{1}{6}\hx_1^4+\frac{1}{12}\hx_1^2\hx_\perp^2
\\
\frac{u_TB^{22}(\bfDelta)}{|\bfDelta|}=\frac{3}{2}-\frac{3}{2}\hx_1^2
+\frac{3}{4}\hx_\perp^2
-\frac{4}{3}\hx_2^2-\frac{1}{6}\hx_1^2\hx_2^2
+\frac{1}{12}\hx_\perp^2\hx_2^2
\end{cases}
\label{4.8}
\eeq
where $\Delta_\perp^2=\Delta_2^2+\Delta_3^2$ and superscripts 2 indicate squares. 
Analyzing Eqn. (\ref{4.8}) in function of
$\hx_1$ first for $\hx_2=0$ and then for $\hx_2=\hx_\perp$, it is possible to show
that $B^{ij}$ is always positive defined, as required.

\vskip 10pt
\noindent{\it Two-dimensional structures}
\vskip 5pt
Suppose that the flow is 2-dimensional, say $\u=\{u_1,0,u_3\}$. 
In this case, the SO(3) decomposition
reduces to an SO(2) one. Keeping again only the lowest order anisotropic correction, we find,
for $\bfDelta=\{\Delta_1,0,\Delta_3\}$:
$$
B^{ij}(\bfDelta)=\frac{|\bfDelta|}{u_T}[(a+3b^{lm}\hx_l\hx_m)\delta^{ij}+\frac{1}{2}(-a+(b^{lm}-c^{lm})
\hx_l\hx_m)\hx^i\hx^j
$$
\beq
-\hx_l[(2b^{li}+c^{li})\hx^j+(2b^{lj}+c^{lj})\hx^i]+3c^{ij}]
\label{4.9}
\eeq
where $b^{22}=b^{33}=0$ and the traceless condition imposes 
$b^{33}=-b^{11}$, $c^{33}=-c^{11}$.

\vskip 10pt
\noindent{\it Streaks}
\vskip 5pt
Two-dimensional 
streaks along the direction of the mean flow appear to be one of the characteristic 
structures in the viscous sublayer of wall turbulence \cite{pope}. 
Contrary to the three-dimensional case, elongated structures cannot be accommodated at $J=2$ in an
SO(2) decomposition: the resulting noise tensor would not be positive definite. Nonetheless,
a noise expression accounting for such structures can still be determined. 
For instance, it is easy to see that, if the streaks are oriented 
along $x_1$ and the flow is two-dimensional in the $x_1x_3$ plane,
an appropriate expression for the noise tensor will be
\beq
B^{ij}(\bfDelta)=a\frac{|\Delta_3|}{u_T}\delta^i_1\delta^j_1
\label{4.10}
\eeq
\vskip 10pt

In the non-Gaussian case, Eqn. (\ref{4.0.0}) ceases to be valid, and the random field
correlation profile ceases to be in general a simple product of exponentials and sines
or cosines. Even if we fit the turbulent correlation with an equation like (\ref{4.0.1}),
the random field correlations will not obey that equation, rather, one involving
higher order correlations. This because of the relation, imposed by the well-mixed 
condition, between non-Gaussian $\rho_E$ and nonlinear ${A_\mu}^i$. 
For instance, if we used a bi-Gaussian distribution to model a high kurtosis PDF 
\cite{luhar89,cohen62,baerentsen84}, a double exponential decay of correlations 
would ensue, with the slower decay associated with the
intermittent bursts (see the end of Appendix A) \cite{maurizi00}. 

For large kurtosis, the noise amplitude determines only the correlation times and
lenghts of the fast decaying exponential. 
The simplest approach, in this case, is to renormalize the noise amplitude, 
with respect to the Gaussian case, in order to correct for the longer correlations
produced by the slowly decaying exponential.
At the end of Appendix A, it is shown that, 
in order to have the desired space and time scales for the bursts,
it is necessary to renormalize the noise amplitude
by a factor $\beta=(2/3)k-1$ with $k$ the kurtosis $[$see Eqn. (A2)$]$.

\vskip 20pt
\section*{ V. Non-uniqueness and the antisymmetric sector}
\vskip 5pt
Once the noise tensor and the PDF $\rho_E$ are fixed and the well-mixed condition is imposed,
the symmetric sector of the velocity correlation is completely determined. 
The non-unique term ${\Xi_\mu}^i$ can be used to fix the structure of the anisotropic sector.
We consider for simplicity the homogeneous case $C^{ij}(x^\mu,\Delta^\mu)=C^{ij}(\Delta^\mu)$.

We discover immediately the following important
fact: not all expressions for the non-unique term ${\Xi_\mu}^i$, and consequently for
${A_\mu}^i$, lead to a statistically realizable $C^{ij}(\Delta^\mu)$.
This is a different face of the problem of local existence for the solutions of Eqn. (\ref{2.8}).
Consider as a first example:
${\Xi_j}^i={\epsilon^{i2}}_ju^2$. A contribution $\Delta C^{11}=
{\epsilon^{12}}_3R^{12}\d x^3$ is then
added to $C^{11}(\d\x)$, with $\d\x=\{0,0,\d x^3\}$,
that has the inadmissible symmetry $\Delta C^{ij}(\d\x)=-\Delta C^{ji}(-\d\x)$. The 
second example is
${\Xi_0}^1=u^2$, ${\Xi_0}^2=-u^1$; in this case we find a contribution $R^{12}\d t$ to
$C^{12}(\d t)$ with the inadmissible symmetry $\Delta C^{12}(\d t)=-\Delta C^{21}(\d t)$.

We seek a form of ${\Xi_\mu}^i$ satisfying all the required symmetries, but still sufficiently 
general to describe most geometric structures one may think of. In analogy with the case
of the noise tensor, this can be done in the frame of an SO(3) expansion starting from $C_A$, 
the antisymmetric component of the correlation [see Eqn. (\ref{3.2})].

As with the noise, the non-unique term ${\Xi_\mu}^i$ can be determined in unique way from $C_A$ in
the case of Gaussian $\u$, fitting the turbulent correlations with exponentials multiplying  
sines or cosines; in this case, ${\Xi_\mu}^i$ will depend linearly on $\u$.
Repeating with $C_A$ the steps followed to obtain the noise tensor in Eqn. (\ref{4.0.3}),
we obtain:
\beq
C_A^{ki}=\frac{1}{2}[{c_j}^iR^{jk}-{c_j}^kR^{ji}]\d s
\label{5.0.1}
\eeq
It turns out that the appropriate quantity on which to carry on the SO(3) expansion is 
not $C_A^{ij}$, rather:
\beq 
S_{ik}S_{jl}C_A^{kl}=\frac{1}{2}[{c_i}^kS_{kj}-{c_j}^kS_{ki}]\d s
=r^\mu{\xi^l}_\mu\epsilon_{lij}\d s+...
\label{5.0.2}
\eeq
where $S_{ij}=(R^{-1})_{ij}$ and the terms in the expansion indicated in the formula are
antisymmetric spherical tensors of order $J=0,1,2$ [see Eqn. (B6)]. Isolating in 
Eqn. (\ref{4.0.0}) the contribution from ${\Xi_\mu}^i$ and using Eqns. (\ref{5.0.1}) and
(\ref{5.0.2}), we obtain therefore for the non-unique term:
\beq
{\Xi_\mu}^i=\rho_E{\xi^l}_\mu\epsilon_{lmj}R^{mi}u^j+...
\label{5.1}
\eeq
and it is immediate to check that the divergence free condition 
$\partial_{u^i}{\Xi_\mu}^i$ $[$the second of Eqn. (\ref{3.7})$]$ is satisfied.

We notice that, had we carried on the expansion directly on $C_A^{ij}$, the tensor $R^{ij}$
entering Eqn. (\ref{5.1}), and consequently the inverse $S_{ij}$ entering $\rho_E$ in 
that formula would have been substituted by the identity matrix $\delta_{ij}$. The contribution
${\Xi_\mu}^i/\rho_E$ to the drift (here $\rho_E$ contains the right $S_{ij}$!) would have
produced therefore explosive behaviors ($\ex^{\alpha|\u|^2}$, $\alpha>0$) in some
direction of $\u$. What happens is that linearity of ${A_\mu}^i$ together with antisymmetry
of $C_A^{ij}$, impose the property $C_A^{ij}=\tilde c_{lm}R^{li}R^{mj}$ with $\tilde c_{lm}$ 
antisymmetric, 
and, keeping only the first terms in the SO(3) expansion for $C_A^{ij}$ would cause loosing
this property.

We still need to enforce incompressibility, i.e. the zero trace condition
$\Xi_i^i=0$ $[$the first of Eqn. (\ref{3.7})$]$.
The fact that the SO(3) 
expansion is carried on $S_{ik}S_{jl}C_A^{kl}$,
rather than on $C_A^{ij}$, will lead to mixing of harmonics of different order $J$ [compare
with the case of the noise tensor and Eqns. (B3-B4)].
It is convenient to separate the antisymmetric part of
${\xi^l}_\mu$:
\beq
{\xi^l}_m=\bar\xi^l_m+{\epsilon^l}_{mk}\zeta^k
\label{5.3}
\eeq
The zero trace condition becomes:
\beq
\Xi^i_i=\rho_E\{\bar\xi^k_iR^{im}\epsilon_{klm}u^l+[R^i_i\zeta_l-R^i_l\zeta_i]u^l\}=0
\label{5.4}
\eeq
which must be satisfied for any $u^l$. This leads to the relation between $\zeta^k$ and
$\bar\xi^l_m$:
\beq
(R^j_l-R^k_k\delta^j_l)\zeta_j=\frac{1}{2}\epsilon_{klm}[\bar\xi R-R\bar\xi]^{km}
\label{5.5}
\eeq
The presence of the commutator $[\bar\xi R-R\bar\xi]^{km}$ suggests that we should
work in the diagonal system for $R^{ij}$. It becomes easy in this way to separate the
part of $\bar\xi^i_j$ which anticommutes with $R^{ij}$, which is simply the part out of
diagonal.
Solution of Eqn. (\ref{5.5}) gives then in the diagonal coordinate system, after little
algebra:
\beq
\zeta_1=\bar\xi_{23}\frac{R_{22}-R_{33}}{R_{22}+R_{33}},
\qquad
\zeta_2=\bar\xi_{13}\frac{R_{33}-R_{11}}{R_{11}+R_{33}},
\qquad
\zeta_3=\bar\xi_{12}\frac{R_{11}-R_{22}}{R_{11}+R_{22}}.
\label{5.6}
\eeq
We are now in the position to determine the effect 
of the various components of ${\Xi_i}^j$,
on ${A_i}^j$ and on the velocity correlations.
\vskip 10pt
\noindent{\it Time component}
\vskip 5pt
It turns out that the $\mu=0$ component of Eqn. (\ref{5.1}) is associated with a combination of
rotation and strain of the velocity, as time passes, at any given position $\x$; this is the
Eulerian version of the mean velocity rotation along Lagrangian trajectories discussed in
\cite{sawford99}.
Working in the diagonal coordinate system for $R^{ij}$,
the contribution from ${\xi^3}_0$ will be, for instance:
\beq
{\Xi_0}^1=\rho_E{\xi^3}_0R^{11}u^2,
\qquad
{\Xi_0}^2=-\rho_E{\xi^3}_0R^{22}u^1
\label{5.7}
\eeq
This turns into a pure rotation if $R^{ij}\propto\delta^{ij}$.

From the point of view of SO(3), this is the $J=1$ contribution to the second of Eqn. (B6), 
which is trivially symmetric in space and antisymmetric in time.
\vskip 10pt
\noindent{\it Space component: diagonal part}
\vskip 5pt
Also this component is associated with a combination of rotation and strain of the velocity, 
this time, as one moves at fixed time from one space point to another. Focusing e.g. on the
contribution from $\bar\xi^2_2$, we find in the diagonal system for $R^{ij}$:
\beq
{\Xi_2}^1=-\rho_E\bar\xi^2_2R^{11}u^3,
\qquad
{\Xi_2}^3=\rho_E\bar\xi^2_2R^{33}u^1
\label{5.8}
\eeq
In the case $R^{ij}\propto\delta^{ij}$, this becomes a pure rotation in the plane $x^1x^3$ as one
moves in the $x^2$ direction. The diagonal component of the non-unique spatial term is the one
associated with the presence of helicity $H$ in the turbulent field. Indicating with 
$\omega^i={\epsilon^{ij}}_k\partial_ju^k$ the vorticity, we can write
\beq
H=\langle u_i\omega^i\rangle=\langle u_i\langle\omega^i|\u\rangle\rangle=
{\epsilon^{ij}}_k\langle u_i{A_j}^k\rangle.
\label{5.9}
\eeq
Substituting the various contributions to ${A_j}^k$ in the above formula, we see that the only
terms giving non-zero result are the diagonal ones in $\bar\xi$. Working in the diagonal 
coordinate system, we obtain then:
\beq
H=2[\bar\xi_{11}R_{22}R_{33}+\bar\xi_{22}R_{11}R_{33}+\bar\xi_{33}R_{11}R_{22}]
\label{5.10}
\eeq
From the point of view of SO(3), this is a combination of $J=2$ contributions from the
second of Eqn. (B6)  (zero trace part of $\xi$) and $J=0$ contributions from the first 
of the same equation.

\begin{figure}[hbtp]\centering
\centerline{
\psfig{figure=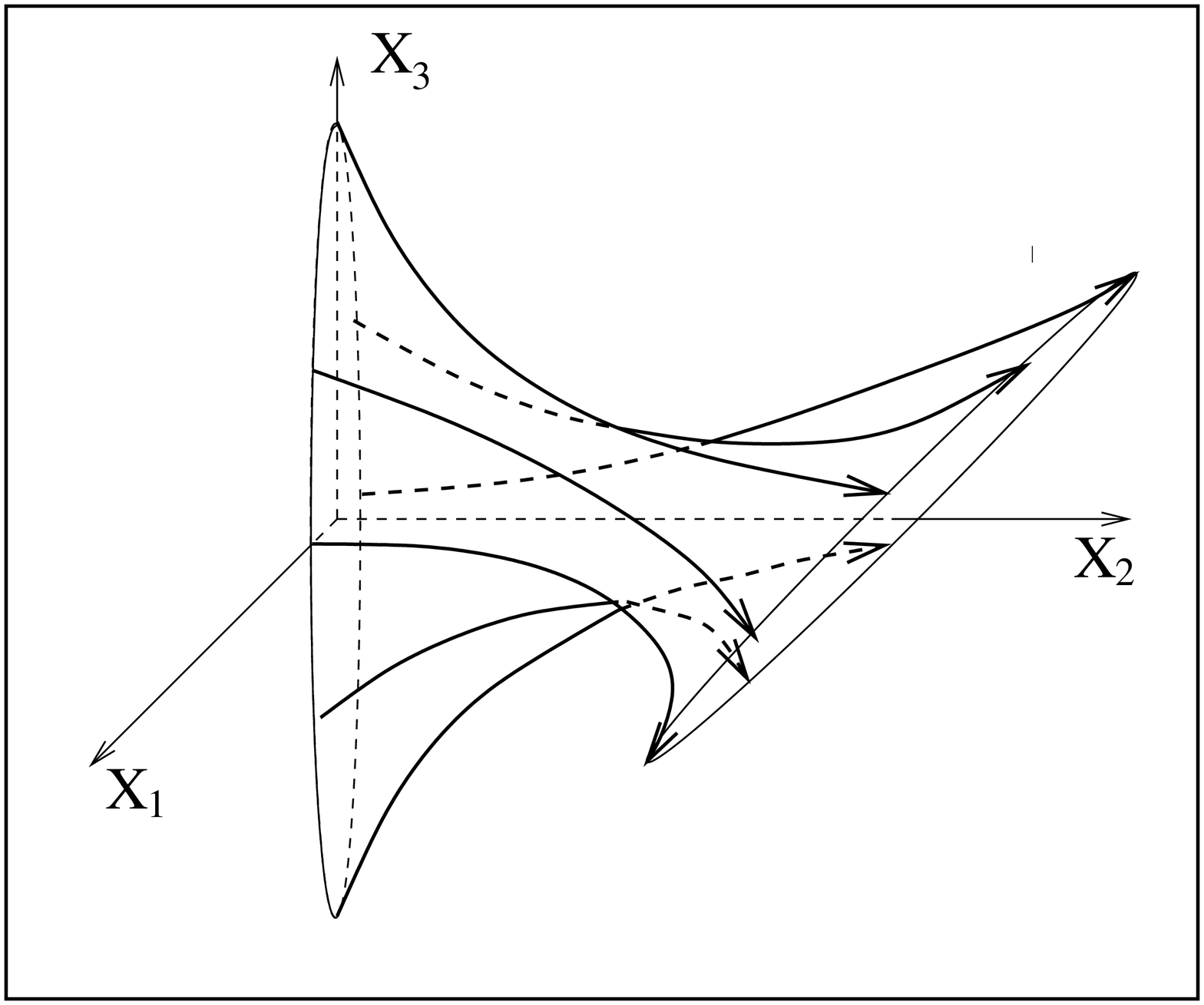,height=6.cm,angle=0.}
}
\caption{Sketch of the velocity lines in a coherent structure characterized by a non-zero 
value of $\bar\xi_1^3$ (part out of diagonal of the non-unique spatial term). The velocity
components in the $13$ plane are arranged along strain line with expanding and compressing
directions respectively along $x_1$ and $x_3$.
}
\label{fig1.fig}
\end{figure}

\vskip 10pt
\noindent{\it Space component: part out of diagonal}
\vskip 5pt
The effect of this component is illustrated, for the contribution from $\bar\xi^3_1$, in 
Figure \ref{fig1.fig}.
This effect consists of a strain of the velocity components as one moves in the 
direction $2$. We give in the equation below the non-zero matrix elements of ${\Xi_k}^i$
corresponding to $\bar\xi^3_1$ (components still evaluated in the diagonal coordinate
system).
\beq
\begin{cases}
{\Xi_1}^1=2\rho_E\bar\xi_1^3\frac{R_{11}R_{33}}{R_{11}+R_{33}}u_2,
\qquad\qquad
{\Xi_1}^2=-\rho_E\bar\xi_1^3R_{22}u_1,
\\
{\Xi_2}^1=2\rho_E\bar\xi_1^3\frac{R_{11}(R_{11}-R_{33})}{R_{11}+R_{33}}u_1,
\qquad
{\Xi_2}^3=2\rho_E\bar\xi_1^3\frac{R_{33}(R_{11}-R_{33})}{R_{11}+R_{33}}u_3,
\\
{\Xi_3}^2=\rho_E\bar\xi_1^3R_{22}u_3,
\qquad\qquad\qquad
{\Xi_3}^3=-2\rho_E\bar\xi_1^3\frac{R_{11}R_{33}}{R_{11}+R_{33}}u_2.
\end{cases}
\label{5.11}
\eeq
From the point of view of SO(3), this is a combination of $J=1$ components from the 
third of Eqn. (B6) (the $\zeta$ piece) and $J=2$ components from the second of the same 
equation (the out of diagonal $\xi$ piece). This
is the case in which the incompressibility condition needs, in order to be enforced,
consideration of spherical tensors of different order $J$.
\vskip 5pt
It is possible to see that, in the non-Gaussian case, all the results obtained 
starting from Eqn. (\ref{5.1}) can be recovered substituting in that equation 
$\rho_E$ with a Gaussian PDF $\rho_G$ with identical $S^{ij}$. Notice that, if a bi-Gaussian
is used to model $\rho_E$, the ratio $\rho_G/\rho_E$ entering the contribution to the
drift will decay like a Gaussian for large $u$, when the slowly decaying Gaussian
entering $\rho_E$, which decays slower than $\rho_G$ as well [see Eqns. (A1) (A2)
and (A4)] become dominant.
However, as in the case of the symmetric sector 
(see discussion at the end of the previous section), the correspondence between 
drift and second order correlations ceases to be unique as Eqn. (\ref{4.0.0}) becomes nonlinear
and Eqn. (\ref{4.0.1}) begins to involve higher order velocity correlations.

\vskip 20pt
\section*{VI. Derivation of Markovian Lagrangian models}
\vskip 5pt
Knowing the form of the tensors
${A_\mu}^i$ and $\langle\d w^i\d w^j\rangle$, allows the derivation of Lagrangian stochastic 
models.
This is done most naturally setting in Eqn. (\ref{2.2}) 
$\d x^\mu=\{\d t,\v\d t\}$, where $\v$ is the
particle velocity.  This entails a Markovian assumption on the Lagrangian 
statistics, whose validity will be checked in the next two sections, although  it  is
not very different from the one used in standard Lagrangian models.

\vskip 10pt
\noindent{\it Passive tracers}
\vskip 5pt
Let us write explicitly the Langevin and Fokker-Planck equations associated with our Lagrangian
model, considering first the simpler case of a passive tracer $\d x^\mu=\{\d t,\u\d t\}$
where $\u(t)$ identifies the fluid velocity sampled by the moving particle:
\beq
\d u^i\equiv\dot u^i\d t=(\u\cdot{\bf A}^i+A_0^{\ i})\d t+\d w^i
\label{6.1}
\eeq
\beq
(\partial_t+\u\cdot\nabla)\rho_L+\partial_{u^i}[(\u\cdot{\bf A}^i+A_0^{\ i})\rho_L]=
\frac{1}{2}\partial_{u^i}\partial_{u^j}({\cal B}^{ij}\rho_L)
\label{6.2}
\eeq
where ${\cal B}^{ij}=\frac{\d}{\d \Delta^0}\langle\Delta w^i\Delta w^j\rangle
=\frac{u_T^2}{\tau_E}[B_t^{ij}(1)+B^{ij}(\u)]$ [see Eqn. (\ref{4.1})] and $\rho_L(\u,\x,t)$ 
is the PDF of finding a Lagrangian tracer at $\x$ with velocity $\u$.

Exactly as in the Thomson-87 approach, in the Gaussian case, 
the contribution to the drift $\u\cdot{\bf A}^i+A_0^{\ i}$ from
turbulence non-homogeneity is at most quadratic in $\u$, 
with the quadratic terms produced by $\u\cdot({\bf\Phi}^i+{\bf\Psi}^i)$. 
However, disregarding the non-unique terms, the form 
of this contributions differs from the one discussed in 
\cite{thomson87} and \cite{sawford99}.

Also the non-unique contribution $\u\cdot{\bf \Xi}^i$, to lowest order in the SO(3) 
expansion, is at most quadratic in $\u$, with the quadratic piece associated with the 
space component $\u\cdot{\bf\Xi}^i$. The observation in \cite{reynolds02a} that 
helical contributions in Lagrangian stochastic models must be quadratic in the 
velocity is thus confirmed. 

Notice that, from the relation $\bfDelta=\u\Delta^0$, 
higher orders in the SO(3) expansion correspond in the Lagrangian model 
to higher order polynomials in $\u$
contributing to ${\cal B}^{ij}$, $\u\cdot\bar{\bf A}^i+{\bar A_0}^i$ and 
$\u\cdot{\bf\Xi}^i+{\Xi_0}^i$. Conversely, at the random field level, 
independently of the order in the SO(3) expansion (and for Gaussian 
statistics), the drift terms are at most linear in $\u$.

The important feature of the model described by Eqns. (\ref{6.1}-\ref{6.2}) is that
the well-mixed condition imposed on the random field, encoded in Eqns. (\ref{3.3}-\ref{3.5})
and (\ref{3.13}) [Eqns. (\ref{3.15}-\ref{3.17}) in the Gaussian case], translates 
automatically into an identical condition on the trajectories. In the incompressible 
case considered here, this condition is equivalent to the ergodic property
$\rho_L(\u|\x,t)=\rho_E(\u,\x,t)$, which will be shown to hold, in the next section,
right thanks to the condition $A^i_i=0$. This property implies trivially that
Eulerian averages $\langle\ \rangle$ and averages along trajectories $\langle\ \rangle_\smalL$
coincide.

At this point, the model described by Eqns. (\ref{6.1}-\ref{6.2}), is undistinguishable
from a model derived through the Thomson-87 technique starting from the same PDF,
the only difference being in the form of the noise term. In its simplest form, the 
noise tensor ${\cal B}^{ij}$ is isotropic, and is obtained by setting 
$\bfDelta=\u\Delta^0$ in Eqn. (\ref{4.5}) and deriving with respect to $\Delta^0$:
\beq
{\cal B}^{ij}=
\frac{2u_T^2}{\tau_E}[\delta^{ij}
+\frac{a|\u|}{u_T}(\delta^{ij}-\frac{u^iu^j}{3|\u|^2})].
\label{6.2.1}
\eeq
This expression must be compared with the one in the Thomson-87 approach:
${\cal B}^{ij}=\delta^{ij}C_0\flux$. 
(As a technical aside, notice that
we started in section II with an additive noise and we have
arrived here at a multiplicative noise term, which is automatically intended, in the
approach that we have followed, in the It\^o sense \cite{gardiner}). 

The difference in the analytical expressions
underlies a difference in physical interpretation: while in the Thomson-87 
technique, ${\cal B}^{ij}\d t$ is precisely the Lagrangian time structure 
function for inertial time separation, in our approach, it is a non-universal
quantity whose form is determined in function of the large scale turbulence 
geometry. In the Thomson-87 approach, the time scale is fixed by the viscous
dissipation $\flux$, which fixes the expression for the Lagrangian correlation
time 
\beq
\tau_L=\frac{2u_T^2}{C_0\flux}.
\label{6.2.2}
\eeq
In our approach, the time scale is fixed directly by $\tau_E$. 
To be precise, the association between $\tau_L$ and $\flux$ in the Thomson-87 
model, is strictly valid only in the Gaussian case \cite{maurizi00}.
Also in our approach, however,
non-Gaussian statistics leads to $\tau_E$ not being directly associated with the
Eulerian time scales, but only with the fast part of the correlation decay
(see end of Appendix A). 

The two approaches depend on dimensionless constants, $C_0$ and $a$, which can
be related in semi-quantitative way.
As discussed in correspondence to Eqn. (\ref{4.5}), the parameter $a$ identifies
a characteristic length for the random field $l_u=u_T\tau_E/a$, which, at least
in the Gaussian case, corresponds to the integral lenght of the turbulence.
Substiting the estimate for the viscous dissipation from our model $\flux\sim u_T^3/l_u$ 
into Eqn. (\ref{6.2.2}) and setting $\tau_E\sim\tau_L$, we obtain:
$$
a\sim C_0^{-1}
$$
This tells us that the Thomson-87 model cannot be recovered
from Eqn. (\ref{6.2.1}), for finite $C_0$, by setting simultaneously
$a=0$ and $\frac{2u_T^2}{\tau_E}=C_0\flux$. The equivalent limits $a\to 0$ and
$C_0\to\infty$ correspond to the regime of  $\tau_E$
much shorter than the eddy rotation time corresponding to the Kraichnan model
\cite{kraichnan94}.
The way in which this limit is carried out, however, is different in the two approaches:
in ours, it is the turbulence integral scale $l_u$ that is sent to infinity 
[see comment after Eqn. (\ref{4.5})]; in the standard approach, the diffusive
limit $C_0\to\infty$ corresponds directly to the one $\tau_E\to 0$.

\vskip 10pt
\noindent{\it Solid particles}
\vskip 5pt
Let us pass to the analysis of the solid particle case.
The solid particle dynamics obeys an equation, which in quite general
form (neglecting memory effects associated with the Basset force) can be written as:
\beq
\dot\v={\bf F}(\v,\u) 
\label{6.3}
\eeq
A general form was derived by \cite{maxey83}, who neglect lift effects 
\cite{saffman64}. These equations are all derived in the limit of particle
diameter $d$ small with respect to the scales of the flow, and
low particle Reynolds number $Re_p=d|{\bf u-v}|/\nu$, where $\nu$ is the kinematic viscosity.
We will consider Eqn. (\ref{6.3}) in simplified form by accounting only for linear Stokes drag
and gravity:
\beq 
\dot{\bf v}=\frac{\u-\v+\v_G}{\tau_S}
\label{6.4}
\eeq
where $\tau_S=1/18\ \gamma d^2/\nu$,  with $\gamma$ the density of the particle relative 
to that of the fluid, is the Stokes time, and $\v_G$ is the particle terminal velocity 
in a uniform force field and a quiescent fluid. In the case of gravity: $\v_G=\tau_S{\bf g}$ with ${\bf g}$ 
the gravitational acceleration. More in general, $\v_G$, may account for body forces like the 
effect of the Saffman lift \cite{saffman64}.

The analogue of Eqn. (\ref{6.1}), in the solid particle case will 
be obtained putting in Eqn. (\ref{2.2})
$\d x^{\mu}=\{\d t,\v \d t \}$, with $\v$ the solid particle velocity: 
\beq
\d u^i\equiv\dot u^i\d t=(\v\cdot{\bf A}^i+A_0^{\ i})\d t+\d w^i
\label{6.5}
\eeq
where now $\u(t)$ is the fluid velocity sampled by the solid particle and $\langle\d w^i\d w^j 
\rangle={\cal B}^{ij}\d t=\frac{u_T^2}{\tau_E}[B_t^{ij}(1)+B^{ij}(\v)]\d t$. Notice that 
the drift tensor $A_{\mu}^{\ i}$ still depends on $\u$, while $\langle\d w^i\d w^j 
\rangle$ depends only on $\v$.
The Lagrangian PDF $\rho_L(\u,\v,\x,t)$ will obey the Fokker-Planck equation:
\beq
(\partial_t+\v\cdot\nabla)\rho_L+\partial_{v^i}(F^i\rho_L)
+\partial_{u^i}[(\v\cdot{\bf A}^i+A_0^{\ i})\rho_L]=
\frac{1}{2}\partial_{u^i}\partial_{u^j}({\cal B}^{ij}\rho_L)
\label{6.6}
\eeq
Equations (\ref{6.3}), (\ref{6.5}) and (\ref{6.6}) are in the standard form for a ''two-fluid''
Lagrangian model for solid particle transport, i.e. a model in which the fluid and solid 
phase are taken into
account at the same time and are treated on the same footing.
All problems in the fluid limit, present in models in which the separation of fluid and solid 
particle trajectories was considered without accounting for the geometry of the process 
\cite{olla02} are clearly avoided: when inertia and gravity are sent to zero, the fluid case
described by Eqns. (\ref{6.1}-\ref{6.2}) is automatically recovered.

We can easily estimate the turbophoretic drift, i.e. the component of
particle transport
due to the turbulence intensity gradient \cite{caporaloni75,reeks83}.
Multiplying Eqn. (\ref{6.6})
by $v^i$ and integrating in
$\v$ and $\u$ we obtain at stationary state and for uniform concentration
$\partial_j\langle
v^iv^j\rangle_\smalL=\langle F^i(\v,\u)\rangle_\smalL$, which,
for linear $F^i$, can be inverted to obtain $\langle v^i\rangle_\smalL$ 
[subscript $L$ indicates that we are averaging over the Lagrangian PDF $\rho_L(\u,\v,\x,t)$].
In the Stokesian case described by Eqn. (\ref{6.6}),
and small Stokes number $St=\tau_S/\tau_E$, we can approximate $\langle v^iv^j\rangle_L=R^{ij}$,
and we obtain:
\beq
\langle v^i\rangle_\smalL=-\tau_S\partial_jR^{ij}+v_G^i
\label{6.7}
\eeq
We can try to understand Eqns. (\ref{6.5}-\ref{6.7}) from the point of view of a model
satisfying the well mixed condition. The particle flow, due to the effect of inertia,
is compressible and preferential concentration phenomena are known to occur \cite{wang93}. 
Therefore, we do not expect in general the ergodic property
$\rho_L(\u,\v|\x,t)=\rho_E(\v,\u,\x,t)$ to be satisfied. 
Turbophoresis provides the simplest illustration of this phenomenon.
Averaging Eqn. (\ref{6.4}) over $\rho_L(\u,\v,\x,t)$ and combining with Eqn. (\ref{6.7}), 
we obtain in fact the relation:
$$
\langle u^i\rangle_\smalL=-\tau_S\partial_jR^{ij}\ne \langle u^i\rangle=0
$$
i.e., in  inhomogeneous turbulence conditions, Eulerian and Lagrangian averages give 
different results.

For these reasons, in the Thomson-87 approach, a two-fluid solid particle transport model, 
would require knowledge, in some reference situation, of the Lagrangian PDF 
$\rho_L(\u,\v,\x,t)$, meaning that informations must be available 
on both the mean particle concentration
$\theta(\x,t)=\int\d^3u\d^3v\rho_L(\u,\v,\x,t)$ and the conditional PDF (the
PDF along a single particle trajectory) $\rho_L(\u,\v|\x,t)$. Notice that this may imply
substituting Eqn. (\ref{6.3}) or (\ref{6.4}) with a model equation whose coefficient
are determined by the well mixed condition.  Actually, one-fluid 
models exist, in which only the particle phase is considered and only the 
PDF $\rho_L(\v|\x,t)$ has to be known, while $\theta$ is obtained from the
continuity equation $\partial_t\theta+\nabla\cdot (\langle\v\rangle_\smalL\theta)=0$
\cite{reynolds99}.

In our approach, knowledge of $\rho_E$ is sufficient to determine the form of the
equation for $\u$ (\ref{6.5}) (the one for $\v$ is unchanged) without any assumption 
on the form of $\rho_L(\u,\v,\x,t)$. This will be shown in Section IX to produce
automatically the correct form of the Lagrangian correlations 
for the fluid velocity along solid particle trajectories, accounting for the
effect of inertia and trajectory crossing \cite{reeks77}.
Notice that, imposing the well mixed condition on 
$\bar\rho_L(\u|\x,t)=\int\d^3v\rho_L(\u,\v|\x,t)$,
within a simplifying ergodic hypothesis $\bar\rho_L=\rho_E$,
is not sufficient to obtain these correct behaviors;
the anisotropic renormalization of the correlation times 
may be accounted for only by ad-hoc modification of the
expression for the noise tensor Eqn. (\ref{1.3}) 
\cite{sawford91}.

\vskip 20pt
\section*{ VII. Lagrangian one-time statistics and ergodic properties}
\vskip 5pt
As discussed at the end of Section II, we can
imagine Eqn. (\ref{2.2}) as giving the local behaviour of a random velocity field
whose restriction to straight lines in space-time are Markovian processes. This allowed to
have a random velocity field with Eulerian correlations in time and space, which are both well
defined and easy to calculate. Unfortunately, unless the $a\to 0$ limit of the Kraichnan
model is considered \cite{kraichnan94}, 
it is not possible to hypothesize at the same 
time a Markovian behaviour along trajectories. In consequence of this, the Lagrangian 
statistics becomes a complicated business.

However, it turns out that different statistical quantities are affected by the presence of memory
in qualitative different ways and there are situations in which Markovianization of the
trajectories becomes appropriate.  Let us try to understand what happens in detail.

The central quantity one needs for a description of Lagrangian statistics are conditional 
probabilities in the form 
\beq
\rho_L(\X_0|\X_1...\X_n)
\label{7.1}
\eeq
where $\X_k\equiv\X(t_k)\equiv\{\u(\x(t_k),t_k),\x(t_k)\}$, $k=0,...n$.
Let us consider for now the simplest case of a passive tracer.
Such conditional probabilities could be obtained ideally by carrying on a Montecarlo of
trajectories originating from $\{t_n,\X_n\}$ and sampling the particle positions and 
velocities at $t=t_k$, $k=n-1,...1$. Let us focus on the case $n=1$, which presents 
already all the difficulties due to memory. Actually, the transition probability
$\rho(\X_0|\X_1)$ gives precisely the evolution of a cloud of tracers from an 
instantaneous release, i.e. a puff; from the same transition probability, 
also the Lagrangian correlation time could be determined and the calculation will be
illustrated in the next section. 
Suppose we have a set of trajectories starting at time $t_1$ with initial condition $\X_1$,
whose form is known up to time $t$. This allows us to reconstruct $\rho_L(\X(t)|\X_1)$.
The conditional probability at the instant $t+\d t$ will be given by the formula:
$$
\rho_L(\X(t+\d t)|\X_1)
$$
\beq
=\int\d^6X(t)\rho_L(\X(t+\d t)|\X(t),\X_1)
\rho_L(\X(t)|\X_1)
\label{7.2}
\eeq
which corresponds to first summing all the trajectories going from 
$\{t_1,\X_1\}$ to $\{t+\d t,\X(t+\d t)\}$ 
passing through $\{t,\X(t)\}$, and then summing over 
$\X(t)$. Now, to determine $\rho(\X(t+\d t)|\X(t),\X_1)$,
we could average first on the part of the trajectories going from 
$\{t_1,\X_1\}$ to $\{t,\X(t)\}$ and then on that going from 
$\{t,\X(t)\}$ to $\{t+\d t,\X(t+\d t)\}$.
From the point of view of a Montecarlo, this means that we can consider an ensemble of 
fictitious trajectories whose dynamics is only conditioned to the initial condition 
$\{t_1,\X_1\}$ and to the current position $\{t,\X(t)\}$.

We thus reach the not so obvious conclusion that, to determine the evolution of a PDF with 
conditions at $n$ previous
instants, we need to study a dynamics conditioned to $n+1$ instants, but we do
not need the whole trajectory history.  Because of this, if we are interested in a 1-time PDF, 
Markovianization of the dynamics will be an appropriate procedure.

This fact allows us to verify analytically that the 1-point velocity PDF sampled by a passive
tracer coincides with the Eulerian PDF; in other words the ergodic property one expects 
from incompressibility is satisfied. 

In the case of a passive tracer, $\d x^\mu=\{\d t,\u\d t\}$
where $\u(t)$ identifies the fluid velocity sampled by the moving particle, and Eqns. (\ref{6.1})
and (\ref{6.2}) will be the Langevin and Fokker-Planck equations associated with the Markovianized 
dynamics.
From incompressibility [see Eqn. (\ref{2.6})], and from the properties
of the drift components $\bar A$, $\Phi$ and $\Psi$ $[$see Eqns.
(\ref{3.3}-\ref{3.6})$]$, we can write:
\beq
(\partial_t+\u\cdot\nabla)\rho_L
+\partial_{u^i}\Phi_0^{\ i}
+u^j\partial_{u^i}\Phi_j^{\ i}=
-\partial_{u^i}[(\bar A_j^{\ i}u^j+\bar A_0^{\ i}-\frac{1}{2}{\cal B}^{ik}
\partial_{u^k})\rho_L] 
\label{7.3}
\eeq
Setting $\rho_L=\rho_E$, from Eqns. (\ref{3.4}) and the time component of Eqn.
(\ref{3.5}), Eqn. (\ref{7.3}) reduces to 
$u^i\partial_i\rho_E=-u^j\partial_{u^i}{\Phi_j}^i$, which, from Eqn.
(\ref{3.5}), is an identity. The ergodic property is thus satisfied.

This is not a trivial property. We can easily construct a counter-example
in which the incompressibility property
$\partial_{\Delta^i}\langle\Delta w^i\Delta w^j\rangle=0$ [the second of Eqn. (\ref{2.6})]
is not satisfied and ergodicity is violated. Considering for simplicity
stationary homogeneous turbulence (hence ${\Phi_\mu}^i={\Psi_\mu}^i=0$) and choosing 
${\Xi_\mu}^i=0$, we have in fact, setting in Eqn. (\ref{3.4}) $\d x^\mu=\{\d t,\u\d t\}$:
\beq
\u\cdot{\bf A}^i+{A_0}^i=\frac{1}{2}{\cal B}^{ij}\partial_{u^j}\log\rho_E
\label{7.3.1}
\eeq
while Eqn. (\ref{6.2}) dictates:
\beq
\u\cdot{\bf A}^i+{A_0}^i=\frac{1}{2}{\cal B}^{ij}\partial_{u^j}\log\rho_L+
\frac{1}{2}\partial_{u^j}{\cal B}^{ij}
\label{7.3.2}
\eeq
Combining Eqns. (\ref{7.3.2}) and (\ref{7.3.3}), leads to a differential equation 
for $\rho_L$ which, due to homogeneity of ${\cal B}$ in $|\u|$,
can be integrated along the direction $\hat\u=\u/|\u|$:
\beq
\rho_L(\u)={\rm const}\ \rho_E(\u) \exp\Big\{-\hat u^j
\int\d s\Big[({\cal B}^{-1})_{jl}\partial_{u^k}{\cal B}^{lk}\Big]_{\u=\hat\u s}\Big\}
\label{7.3.3}
\eeq
Taking a noise term not satisfying incompressibility,
e.g. $B^{ij}(\bfDelta)=\frac{2u_T}{\tau_E}|\bfDelta|\delta^{ij}$, we would 
obtain $\rho_L(\u)={\rm const}\, |\u|^{-1}\rho_E(\u)$ and 
ergodicity violation.

\vskip 5pt
We can repeat the calculation to check for departures from ergodicity in the solid particle
case. Ergodicity means in this case that the fluid velocity distribution sampled by the 
solid particle
\beq 
\bar \rho_L(\u|\x)=\int \rho_L(\u,\v|\x) d\v
\label{7.4}
\eeq
coincides with $\rho_E(\u,\x)$. 
In all the Montecarlo simulations that we have carried on, described in detail in Section IX,
we have found that, despite compressibility of the solid particle flow, ergodicity was 
satisfied in isotropic homogeneous conditions. The mechanism seems to be the following.

In homogeneous stationary conditions, the
Fokker-Planck equation for the distribution $\rho_L(\u,\v)$ will read, from Eqn. (\ref{6.6}):
\beq
\partial_{v^i}(F^i\rho_L)+
\partial_{u^i}[(\v\cdot{\bf A}^i+{A_0}^i)\rho_L]=
\frac{1}{2}\partial_{u^i}\partial_{u^j}{\cal B}^{ij}\rho_L
\label{7.6}
\eeq
where ${\cal B}^{ij}={\cal B}^{ij}(\v)$.
Exploiting well-mixed [see Eqn. (\ref{3.4})] and setting from isotropy $\Xi=0$,
this equation can be rewritten in the form:
\beq
\partial_{v^i}(F^i\rho_L)+\partial_{u^i}[\frac{1}{2}{\cal B}^{ij}(\partial_{u^j}\rho_L
-\rho_L\partial_{u^j}\log\rho_E)]=0
\label{7.7}
\eeq
and, integrating in $\d^3v$, we reach the following equation for the deviation from ergodicity:
\beq
\partial_{u^i}[\bar\rho_L(\langle {\cal B}^{ij}|\u\rangle
\partial_{u^j}\log\bar\rho_L/\rho_E
+\partial_{u^j}\langle{\cal B}^{ij}|\u\rangle)]=0
\label{7.9}
\eeq
We see that non-ergodic behaviours are associated with the divergence of the 
average over solid particle trajectories of the velocity structure function:
\beq
\partial_{u^j}\langle{\cal B}^{ij}|\u\rangle=\int\d^3v\partial_{u^j}\rho_L(\v|\u){\cal B}^{ij}(\v)
\label{7.10}
\eeq
In the case of solid particles, for which  $\langle{\cal B}^{ij}|\u\rangle\ne{\cal B}^{ij}(\u)$, 
we would expect in general $\partial_{u^j}\langle{\cal B}^{ij}|\u\rangle\ne 0$.
This turns out not to be true, however, when turbulence is isotropic. Let us show how this 
happens.

We can decompose $\nabla_u\rho_L(\v|\u)$ in spherical vectors depending on $\v$ 
[see Eqns. (B7-B8)]:
\beq
\nabla_u\rho_L(\v|\u)=\rho_{01}\v+\rho_{11}(\u\cdot\v)\v+\rho_{12}\u+...
\label{7.11}
\eeq
where, from isotropy, $\rho_{lk}=\rho_{lk}(|\v|,|\u|)$. Higher harmonics (not indicated)
are by construction orthogonal (see Appendix B). From Eqn. (\ref{4.5}), we see that only the term
\beq 
{\bf h}= \rho_{11}(\u\cdot\v)\u+ \rho_{12}\u
\label{7.12}
\eeq
can contribute to $\partial_{u^j}\langle{\cal B}^{ij}|\u\rangle$.
In order for this contribution to be zero, it is sufficient that the curl with respect to $\v$ of
${\bf h}$ be identically zero:
\beq
[|\v|^{-1}\partial_{|\v|}\rho_{12}-\rho_{11}]\v\times\u=0
\label{7.13}
\eeq
so that ${\bf h}$ can be written in the form of a potential term $\nabla_v g(\u,\v)$, and, 
substituting into Eqn. (\ref{7.10}) and integrating by parts:
$$
\partial_{u^j}\langle{\cal B}^{ij}|\u\rangle=-\int\d^3 vg(\u,\v)\partial_{v^j}{\cal B}^{ij}(\v)=0.
$$
Now, we can obtain Eqn. (\ref{7.13}) simply by imposing the condition, from isotropy:
\beq
\partial_{u^i}\langle|\v|^nv^jv^k|\u\rangle=\partial_{u^j}\langle|\v|^nv^iv^k|\u\rangle,
\label{7.14}
\eeq
with $i\ne j\ne k$ and $n\ge 0$. In fact, writing the averages in explicit form, Eqn. (\ref{7.14}) 
can be shown to be equivalent to:
\beq
\int\d^3 v|\v|^{n+1}v^k[\partial_{v^i}\partial_{u^j}-\partial_{v^i}\partial_{u^j}]\rho_L(\v|\u)
\label{7.15}
\eeq
and again, from orthogonality of the decomposition, only the $\rho_{11}$ and $\rho_{12}$ 
terms in $\nabla_u\rho_L$ could contribute. Hence, exploiting the fact that $\rho_{ij}$ 
depends only on $|\u|$ and $|\v|$, Eqn. (\ref{7.15}) becomes equivalent to 
$$
\int\d^3 v[|\v|^{-1}\partial_{|\v|}\rho_{12}-\rho_{11}]|\v|^{n+1}=0
$$
which implies Eqn. (\ref{7.13}) and satisfaction of the ergodic property.

\vskip 20pt
\section*{ VIII. Two-time statistics and the Lagrangian correlation time}
\vskip 5pt
Explicit determination of the Lagrangian dynamics taking into account memory of an initial 
condition is possible when the $\u(\x,t)$ is isotropic, homogeneous and Gaussian.
It thus becomes possible to estimate the error implied in the Markovianization of the
trajectories. The simplest estimator is the Lagrangian correlation time
\beq
\tau_L=\frac{1}{3u_T^2}\int_0^\infty\d t\langle\u(t)\cdot\u(0)\rangle
\label{8.1}
\eeq
where $\u(t)\equiv\u(\x(t),t)$ and we are considering passive tracers.
As discussed at the start of the previous section, we need an evolution equation for 
the trajectory $\{\u(t),\x(t)\}$, given an initial condition 
at $t=0$ $[$for simplicity, fix $\x(0)=0]$. 
The starting point is the following decomposition for the tracer velocity:
\beq
\u(t+\Delta)=\langle\u(t+\Delta)|\u(t),\x(t);\u(0),0\rangle+\Delta\w
\label{8.2}
\eeq
plus knowledge of the conditional averages:
\beq
\langle\u(t+\Delta)|\u(t),\x(t);\u(0),0\rangle
\quad
{\rm and}
\quad
\langle\Delta\w\Delta\w|\u(t),\x(t);\u(0),0\rangle
\label{8.3}
\eeq
As discussed in detail in Appendix C, these averages can be obtained from the correlation 
between velocities at $\{0,0\}$,
$\{t,\x(t)\}$ and $\{t+\Delta,\x(t+\Delta)\}$. We identify 
correlations between points on a trajectory by:
\beq
\hat C^{ij}(\Delta)=\langle u^i(t)u^j(t+\Delta)\rangle,
\label{8.4}
\eeq
If $\u$ is Gaussian, homogeneous and isotropic, these correlations can be expressed in 
analytical form. The mean rate of fluid velocity change along a generic space-time direction 
$\{1,\V\}$ will in this case take the form: 
\beq
{A_0}^i+V^j{A_j}^i=-\frac{1}{\tau_E}\Big[\Big(1+a\frac{|\V|}{u_T}\Big)u_\perp^i
+\Big(1+a\frac{2|\V|}{3u_T}\Big)u_\parallel^i\Big]
\label{8.5}
\eeq
where $\u_\perp$ and $\u_\parallel$ are the components of $\u$ perpendicular and parallel to
the fixed direction
$\V$, and we have used Eqns. (\ref{3.14}), (\ref{3.15}) and (\ref{4.5}), and the expression
$R^{ij}=u_T^2\delta^{ij}$.

Solving the equations for the correlation function along $\{1,\V\}$:
\beq
\frac{\d}{\d t}\langle u^i(\V t,t)u^j(0,0)\rangle=\langle[{A_0}^i+V^j{A_j}^i]u^j(0,0)\rangle
\label{8.6}
\eeq
and introducing longitudinal and transverse projectors
\beq
\Pi^{ij}_\parallel(\V)=\frac{V^iV^j}{|\V|^2},
\qquad
\Pi^{ij}_\perp(\V)=\delta^{ij}-\Pi^{ij}_\parallel(\V)
\label{8.7}
\eeq
we obtain:
\beq
\langle u^i(\V t,t)u^j(0,0)\rangle=\Pi_\perp^{ij}(\V)C_{V\perp}(t)
+\Pi_\parallel^{ij}(\V)C_{V\parallel}(t)
\label{8.8}
\eeq
where
\beq
C_{V\perp}(t)=u_T^2\exp\Big(-\Big(1+\frac{a|\V|}{u_T}\Big)\frac{t}{\tau_E}\Big)
\label{8.9}
\eeq
and
\beq
C_{V\parallel}(t)=u_T^2\exp\Big(-\Big(1+a\frac{2|\V|}{3u_T}\Big)\frac{t}{\tau_E}\Big)
\label{8.10}
\eeq
We shall need also the inverse $D_{ij}(t_l-t_m)$ of the correlation matrix 
$\langle u^i(t_l)u^j(t_m)\rangle$, $l,m=1,2$; $t_1=0$, $t_2=t$, defined by the relation: 
\beq
\sum_mD_{ij}(t_l-t_m)\langle u^j(t_m)u^k(t_n)\rangle=
\sum_m\langle u^k(l_n)u^j(t_m)\rangle D_{ji}(t_m-t_l)=\delta^k_i\delta_{ln}
\label{8.11}
\eeq
From Eqns. (\ref{8.8}-\ref{8.10}), we find:
\beq
D_{ij}(t_l-t_m)=\Pi^\perp_{ij}(\U)D_\perp(t_l-t_m)+\Pi^\parallel_{ij}(\U)D_\parallel(t_l-t_m)
\label{8.12}
\eeq
where $\U=\u(t)-\u(0)$,
\beq
D_\perp=
\frac{1}{C_{U\perp}^2(0)-C_{U\perp}^2(t)}
\begin{pmatrix}
C_{U\perp}(0)  &  -C_{U\perp}(t)
\\
-C_{U\perp}(t) &   C_{U\perp}(0)
\end{pmatrix}
\label{8.13}
\eeq
and we have similar expression for $D_\parallel$.
At this point, we can obtain from Eqn. (C7) the expression for the average evolution of the
velocity along a trajectory, conditioned to an initial condition at time zero:
$$
\langle u^i(t+\Delta)|\u(t),\x(t);\u(0),0\rangle=[\hat C^{ij}(\Delta)D_{jk}(t)
+\hat C^{ij}(t+\Delta)D_{jk}(0)]u^k(0)
$$
\beq
+[C^{ij}(\Delta)D_{jk}(0)+C^{ij}(t+\Delta)D_{jk}(t)]u^k(t)
\label{8.14}
\eeq
As obvious, memory of the initial condition at time zero is lost when $t\to\infty$.
Notice that, if all points $\{0,0\}$, 
$\{t,\x(t)\}$ and $\{t+\Delta,\x(t+\Delta)\}$ are aligned along the same space-time 
direction $\{1,\V\}$, all the $\Pi_\parallel$ and $\Pi_\perp$ entering the correlation 
functions in Eqn. (\ref{8.12}) will project along or perpendicular the same vector $\V$.
In this case the components of $\u$ parallel and perpendicular
to $\V$ decouple and the indices disappear; for instance:
$$
\langle u_\parallel(t+\Delta)|\u(t),\x(t);\u(0),0\rangle=[C_{V\parallel}(\Delta)D_{V\parallel}(t)
+C_{V\parallel}(t+\Delta)D_{V\parallel}(0)]u_\parallel(0)
$$
\beq
+[C_{V\parallel}(\Delta)D_{V\parallel}(0)+C_{V\parallel}(t+\Delta)D_{V\parallel}(t)]u_\parallel(t)
\label{8.15}
\eeq
and all the correlations have the same decay rate fixed by Eqn. (\ref{8.10}). It is then easy to
show that the first term on the RHS of Eqn. (\ref{8.15}) disappears and we have
\beq
\langle u_\parallel(t+\Delta)|\u(t),\x(t);\u(0),0\rangle=
u_\parallel(t)
\exp\Big(-\Big(1+a\frac{2|\V|}{3u_T}\Big)\frac{\Delta}{\tau_E}\Big)
\label{8.16}
\eeq
Hence $\langle u_\parallel(t+\Delta)|\u(t),\x(t);\u(0),0\rangle=
\langle u_\parallel(t+\Delta)|\u(t),\x(t)\rangle$.
If the trajectory is developing along a straight line, we will recover Markovian statistics as 
required.

\begin{figure}[hbtp]\centering
\centerline{
\psfig{figure=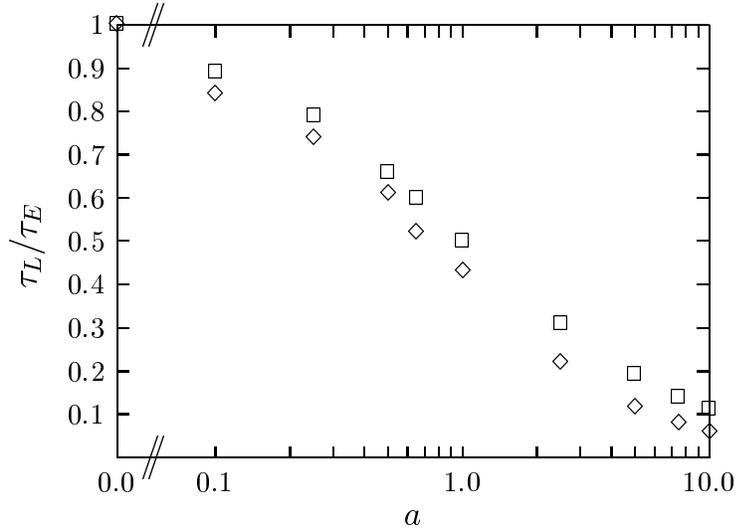,height=7.cm,angle=0.}
}
\caption{Dependence of the Lagrangian correlation time on the ratio $a$ between
eddy life time and eddy rotation time; $\Box$ exact; $\Diamond$ Markovian approximation. 
}
\label{fig2.fig}
\end{figure}

Expressing $\Delta\w$ as the difference between 
$\langle\u(t+\Delta)|\u(t),\x(t);\u(0),0\rangle$ and $\u(t+\Delta)$, 
and substituting into Eqn. (C8), we obtain instead, for the fluctuation term:
$$
\langle\Delta w^i\Delta w^j|\u(t),\x(t);\u(0),0\rangle=
$$
$$
=C^{ij}(0)-D_{kl}(0)[C^{li}(\Delta)C^{kj}(\Delta)+C^{li}(t+\Delta)C^{kj}(t+\Delta)]
$$
\beq
-D_{kl}(t)[C^{li}(\Delta)C^{kj}(t+\Delta)+C^{li}(t+\Delta)C^{kj}(\Delta)]
\label{8.17}
\eeq
In Figure \ref{fig2.fig} we compare the result of a Montecarlo for the Lagrangian correlation time 
Eqn. 
(\ref{8.1}) using the exact dynamics described by Eqns. (\ref{8.2}), (\ref{8.14}) and
(\ref{8.17}), with that obtained from the Markovianized version given by Eqn. (\ref{6.1}). 
As could be guessed, the Markovian approximation becomes exact in the $a=0$ limit, when the 
trajectory, in a correlation time, remains close to the time line $\{1,0\}$. At least in 
this case, the choice given by Eqn. (\ref{2.9}), 
of Markovian statistics along rectilinear cuts in space-time,
is the most appropriate.

\vskip 20pt

\vskip 20pt
\section*{IX. Solid particle transport in homogeneous isotropic turbulence}

\vskip 5pt
Inertia and crossing trajectory effects determine a
substantial change in the statistics of fluid velocities sampled by the solid particle with respect
to that of passive tracer velocities. 

Several authors reserved particular attention to the long time behaviour of 
correlation functions of sampled fluid velocities and long-time particle
diffusion coefficients \cite{csanady63,reeks77,pismen78,nir79}.

Following Csanady \cite{csanady63}, Sawford and Guest \cite{sawford91} kept into
account the effect of gravity produced trajectory 
crossing by a suitable assumption on the renormalization of the correlation
time of the fluid velocity sampled by the falling particle.
Their model was applied to grid-generated turbulence and their results
were found to agree with experimental wind tunnel data.
It is not clear, however, how much this approach can be extended to generic
non-homogeneous and non-stationary turbulent flows, especially in the case
of strong turbulence gradients \cite{reynolds00}. 

Free-flight models \cite{friedlander57} (see also \cite{kallio89} for a brief review) 
are known to make unphysical assumptions about the
velocity that particles assume when they are projected towards the wall from the
buffer and logarithmic regions.
As regards the eddy-interaction model of Kallio and Reeks \cite{kallio89},
this has been shown in \cite{underwood93} not to satisfy the well-mixed condition.
The model described in \cite{underwood93} improved this aspect, but 
without reproducing the build-up of concentration. The issue, to be discussed in the
next section, is the difficulty in isolating near wall solid particle accumulation effects
from spurious concentration build-up from unproper treatment of the well-mixed condition.

A recent advance was obtained in \cite{reynolds99}, but in this
approach a turbophoretic force had to be introduced from the outside, whereas in our approach
the turbophoretic flux results in self-consistent way from the dynamics [see Eqn. (\ref{6.7})].
\vskip 5pt

The central role in the solid particle dispersion is played by
the correlation time $\tau'_L$, of the fluid velocity sampled by the solid
particles. 
In particular, with $\tau'_L(\parallel)$ we indicate the longitudinal effective Lagrangian time, i.e.,
along the direction of gravity, and with $\tau'_L(\perp)$ the transverse effective Lagrangian 
time. 

Let us briefly summarize the main properties of $\tau'_L(i)$.
When gravity is dominant ($v_G \gg u_T$), the correlation function of sampled fluid
velocities decays faster than that of passive tracers and
$\tau'_L (i) < \tau_L$, where $i$ is referred to longitudinal ($\parallel$) or transverse
($\perp$) \cite{csanady63,reeks77}. Taking $v_G=g\tau_S$ with $g$ fixed, we see that $\tau'_L(i)$
decreases from the value $\tau_L$, corresponding to $\tau_S=0$, to zero as
$\tau_S$ increases. Furthermore, the correlation functions do not decay in the
same way in all directions. Due to the continuity effect described in 
\cite{csanady63}, the
decay is slower in the direction of gravity, so that the longitudinal
Lagrangian time $\tau'_L(\parallel)$ is longer than the transverse
one $\tau'_L(\perp)$.

In the inertia dominated case ($v_G \ll u_T$),
the sampled correlation function decays
slower than that of passive tracer velocities and $\tau'_L > \tau_L$. The limit
$\tau_S \to 0$ is the same as above, but now $\tau'_L$ increases with
$\tau_S$ and, in the limit $\tau_S \to \infty$, tends to the Eulerian
time scale $\tau_E$ \cite{reeks77,pismen78}.

We will show shortly how all these effects are automatically 
reproduced in our approach.
\vskip 5pt

We consider a Gaussian homogeneous and stationary isotropic zero-mean random
velocity field.
Thus, the drift term $\bar A_\mu^{\ i}$ is given by Eqn. (\ref{3.15}) with
$S^{ij}=\delta^{ij}/u^2_T$, the PDF is given by Eqn. (\ref{3.14}) and the 
noise tensor is isotropic 
[see Eqn. (\ref{4.5})]. The terms $\Phi^{\ i}_\mu$ and $\Psi^{\ i}_j$ are zero
for homogeneity and the non-unique terms $\Xi_\mu^{\ i}$ are zero for isotropy.

From now on in this section, we rewrite the equations expressing the
velocities $\u$ and $\v$ and time $t$ in units of $u_T$ and $\tau_E$
respectively.
Note that, in this way, $\tau_S$ becomes equivalent to the Stokes number
$St$ \cite{maxey83}, i.e.,
the ratio between the Stokes time and a flow time scale ($\tau_E$ in this case). 
As regards gravity,
it results $v_G=St/Fr$, being $Fr=g\ \tau_E/u_T$ the Froude number related to
the magnitude of the gravity $g$ with respect to turbulence scales \cite{reeks77}.

Under these conditions, Eqn. (\ref{6.5}) for the sampled fluid velocity $\u (t)$
and the associated expression for the noise tensor ${\cal B}^{ij}$ will take 
the simplified form: 
\beq
\begin{cases}
\d u^i=-(1+a\ |\v|)u^i\d t+\frac{a}{6} v^j u_j \hat v^i \d t+dw^i 
\\
\langle dw^i dw^j\rangle =2 \left[ \left( 1+a\ |\v| \right)\delta^{ij}-\frac{a}{3}
|\v|\  \hat v^i \hat v^j \right] dt
\end{cases}
\label{moteq_sfv}
\eeq
with $\hat{\v}=\v/|\v|$ and $\v$ the particle velocity, whose dynamics is given by
Eqn.(\ref{6.4}).
Applying to Eqn. (\ref{moteq_sfv}) the projectors defined in Eqn. (\ref{8.7}),
with ${\bf V}=\v$, it is easily seen that $\d\u$ is split into a longitudinal
and a transverse component, characterized by different values of the drift: $1+2/3\ a
|\v|$ for the fluid velocity component parallel to the particle velocity and
$1+a|\v|$ for the normal component. 
As the role of gravity increases, the symmetry breaking of particle motion due
to the presence of a preferential direction, i.e., the direction of gravity,
involves a separation between the longitudinal and transverse time scale 
(continuity effect).
In the gravity dominated case, $v_G \gg 1$, we have $|\v| \simeq v_G$, with the
result:
\beq
\tau'_L(\parallel) \simeq \frac{3}{2a v_G};
\qquad 
\tau'_L(\perp)\simeq \frac{1}{a v_G};
\qquad
\frac{\tau'_L(\parallel)}{\tau'_L(\perp)}=\frac32 
\label{grav_dominant}
\eeq
It is difficult to obtain an analytical solution for the PDF $\rho_L(\u,\v)$
and the velocity correlation functions $\langle u^i(t) u^j(0) \rangle$
because of the multiplicative noise. For this reason, numerical
simulations by means of Montecarlo technique have been performed to obtain
solutions of Eqns. (\ref{6.4}) and (\ref{moteq_sfv}).
Following Yeung and Pope \cite{yeung89} (see also \cite{sato87}), we choose a
value $\tau_L/\tau_E\thickapprox 0.5$, which, from Figure \ref{fig2.fig}, 
corresponds to $a=0.65$.

As mentioned in Section VII, the ergodic property has been verified to hold also
in the solid particle case.
Both in the case of Gaussian statistics and of an isotropic kurtosis described
by Eqn. (A3) (see Appendix A), with and without gravity, the marginal Lagrangian
PDF $\bar\rho_L(\u)$ defined in Eqn.
(\ref{7.4})  has been found to coincide, to within numerical error, 
with the Eulerian PDF $\rho_E(\u)$.

In the presence of gravity, this means that the average of the
sampled fluid velocity $\langle\u\rangle_L$ coincides with its Eulerian 
counterpart $\langle\u\rangle$, which is zero, and that, therefore,
no renormalization is produced on the value of the terminal velocity
$\v_G$.

Another non trivial result from the numerical simulation is that
the correlation functions of both passive tracers velocities 
($\tau_S=0$ and $v_G=0$) and of sampled fluid velocities appears to
decay exponentially as in the Gaussian case for standard Lagrangian models.

\begin{figure}[hbtp]\centering
\centerline{
\psfig{figure=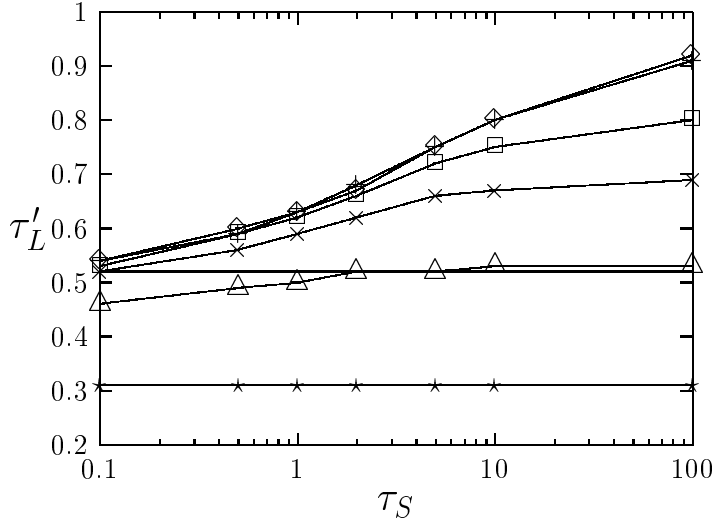,height=7.cm,angle=0.}
}
\caption{Behaviour of $\tau'_L(\perp)$
as a function of the Stokes
time $\tau_S$ for different values of the adimensional terminal velocity $v_G$
(dimensionless units). $\Diamond\ v_G=0$; $+\ v_G=0.1$; $\Box\ v_G=0.5$; $\times\ 
v_G=1$; $\triangle\ v_G=2$; $\star\  v_G=5$. Reference line at $\tau'_L=\tau_L=0.52$. 
}
\label{fig3.fig}
\end{figure}

\begin{figure}[hbtp]\centering
\centerline{
\psfig{figure=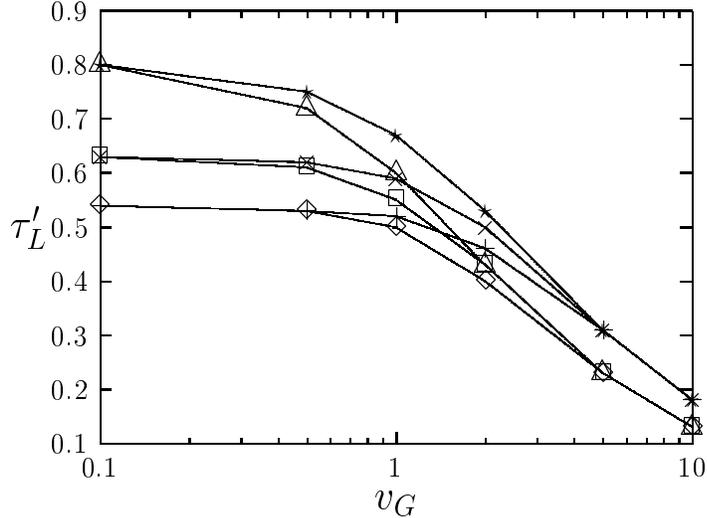,height=7.cm,angle=0.}
}
\caption{Behaviour of $\tau'_L(\parallel)$ and $\tau'_L(\perp)$ as a function of
the terminal velocity $v_G=$ for different values of $\tau_S$ (dimensionless units). 
Longitudinal: $\star\ \tau_S=10$; $\times\ \tau_S=1$; $+\ \tau_S=0.1$.
Transverse:  $\triangle\ \tau_S=10$; $\Box\ \tau_S=1$; $\Diamond\ \tau_S=0.1$.
}
\label{fig4.fig}
\end{figure}

In Figure \ref{fig3.fig} the effective transverse Lagrangian times
$\tau'_L(\perp)$ have been plotted as function of $\tau_S$ ( in units of
$\tau_E$) for different values of
$v_G$ (the Lagrangian time scale of passive tracer has been reported for a
comparison).
In the range $v_G < 1$ (inertia dominant) the curves are increasing from $\tau_L/\tau_E=0.52$
and tend to about $1$ as $\tau_S\rightarrow\infty$ (i.e., the Lagrangian time
tends to the Eulerian time). For $v_G > 1$ (gravity dominant) the curves
loose their dependence on $\tau_S$ and the correlation time is approximately
equal to the eddy crossing time given (always in dimensionless units) by 
$v_G^{-1}$.  This is in agreement with the asymptotic formulae in Eqns.
(\ref{grav_dominant}).
Figure \ref{fig4.fig} shows the behaviour of $\tau'_L(\parallel)$ and
$\tau'_L(\perp)$ as function of $v_G$ for fixed $\tau_S$. As expected, each
curve tends to a constant value in the limit $v_G\to 0$.
The longitudinal times
are always longer than the transverse ones and, in the limit $v_G\to
\infty$, they collapse onto two different curves, whose ratio is $3/2$
as predicted by Eqn. (\ref{grav_dominant}).

An exponential decay of the sampled fluid velocity correlation function allows
easy analytical calculation of the correlation function for $\v$
\cite{pismen78,nir79,meek73}.
The last one reads, for $i=\ \parallel,\perp$:
\beq
C_p^i(t) =\langle v^i(t) v^i(0) \rangle
= C_p^i(0) \left[ 
\ex^{-t/\tau_S}+
\frac{\ex^{-t/\tau'_L(i)}-\ex^{-t/\tau_S}}{1-\tau_S/\tau'_L(i)} \right]
\label{corr_v}
\eeq
where
\beq
 C_p^i(0)=\frac{u^2_T}{1+\tau_S/\tau'_L(i)}
\label{corr_v0}
\eeq
The particle correlation time $\tau_p(i)$ can then be calculated 
$$
\tau_p(i)=\int_0^{\infty} \frac{C_p^{i}(t)}{C_p^{i}(0)}
dt=\tau_S+\tau'_L(i) 
$$
By using Taylor's theorem, the (long-time) diffusion coefficients
$$
\kappa(i)=\frac12 \lim_{t \rightarrow \infty} \frac{1}{t}
\langle \left(x_i(t)-x_i(0)\right)^2 \rangle,
\qquad
i=\ \parallel,\perp
$$
can be expressed in terms of the Lagrangian correlation times for $\v$ as follows
$$
t\gg \tau_p(i): \qquad \kappa(i)=
\tau_p(i) C_p^{i}(0)= \tau'_L(i)
$$
Thus, the diffusion coefficients $\kappa(i)$
will behave exactly as the effective Lagrangian times $\tau'_L(i)$.
The adimensional diffusion coefficient
of passive tracers is simply given by $\kappa=\tau_L=0.52$. Hence, in agreement
with \cite{reeks77}, $\kappa(i)$ will be larger than in the passive scalar case
when inertia is dominant, smaller when gravity is dominant. Furthermore, 
when gravity is dominant, the longitudinal solid particle diffusion 
coefficient will be larger than the transverse one.

\vskip 20pt
\section*{X. Solid particle transport in turbulent channel flow}
\vskip 5pt
We focus in this section on phenomena of accumulation and deposition associated
with the interaction of inertial particles with the inhomogeneity of the flow
and the presence of solid boundaries. Starting from the work
of McLaughlin \cite{mclaughlin89}, the reference situation that is typically
considered, both to identify the main features of 
particle transport, and to test the functionality of transport models, is that
of the turbulent channel flow.  


We have tested our model in its simplest form, with a Gaussian PDF, an isotropic
noise and non-unique term ${\Xi_\mu}^i$ set to zero. We recall that, in this form, the model is
described by Eqns. (\ref{6.4}-\ref{6.5}), with the drift given by Eqns. (\ref{3.3}) and
(\ref{3.15}-\ref{3.17}), and the noise by Eqn. (\ref{4.5}) with $\bfDelta=\v\Delta^0
\equiv\v\Delta t$. As in the homogeneous-isotropic turbulence case, we have set the
free parameter $a=0.65$, corresponding to the value of the
ratio between Lagrangian and Eulerian correlation time 
$\tau_L/\tau_E=0.52$ \cite{yeung89,sato87}.
We adopt standard wall variables identified where necessary with $+$, normalized with
the friction velocity $u_*$ and the reference length and time scales 
$x_2^*=\nu/u_*$ and $\tau_*=x_2^*/u_*$ where $\nu$ is the kinematic viscosity.

For the Lagrangian correlation time we have used the interpolation formula:
\beq
\begin{cases}
\tau_L=7.122+0.5731\ x_2^+-0.00129(x_2^+)^2\quad &x_2^+<140
\\
\tau_L=-19.902+.959\ x_2^+-.00267(x_2^+)^2 &140<x_2^+<180
\end{cases}
\label{kallio}
\eeq
where $x_2$ identifies the cross-stream direction
(we take $x_1$ and $x_3$ respectively in the streamwise and spanwise direction).
For $x_2^+<140$, Eqn. (\ref{kallio}) coincides with the interpolation formula quoted 
in \cite{kallio89}.
Thus, Eulerian time scales span from $\tau_E^{max} \simeq 140$ in
the channel centre to $\tau_E^{min} \simeq 14$ at the walls.

We have 
considered neither the effect of gravity, nor that of Brownian motion. The second may
be important in the case of sub-micrometer particles.
We have included, instead, the contribution from the Saffman lift \cite{saffman64};
indicating as usual with $d$ the particle diameter and $\gamma$ the particle to fluid density
ratio:
$$
v_G=0.39\frac{\tau_S}{\gamma d}\Big|\nu\frac{\partial\bar u_1}{\partial x_2}\Big|^\frac{1}{2}
{\rm sign}\Big(\frac{\partial\bar u_1}{\partial x_2}\Big)
(u_1-v_1),
$$
which is known to contribute to the solid particle dynamics in the range 
$\tau_S\lesssim 10$ \cite{kallio89}.
\vskip 5pt
As input data for the model, we have utilized the 1-point statistics from the 
DNS by Kim, Moin and Moser \cite{kim87}.
The channel width $L_c$ is nearly $360$ wall units and the Reynolds number is of order $3300$,
based on the maximum mean velocity and the channel half-width.

We have carried on Montecarlo simulations with $N_{tot}=10000$
particles, uniformly distributed at the initial time.
In order to obtain, after a suitable time, a stationary concentration profile, 
we have introduced a source of particles at the channel centre to
balance the deposition flux at the walls. 
As in \cite{brooke94}, for these conditions and  for quite
all Stokes times, it seems that a simulation time $T_{sim}$ of about $700$ is
sufficient to achieve a stationary distribution for the solid particles.

\begin{figure}[hbtp]\centering
\centerline{
\psfig{figure=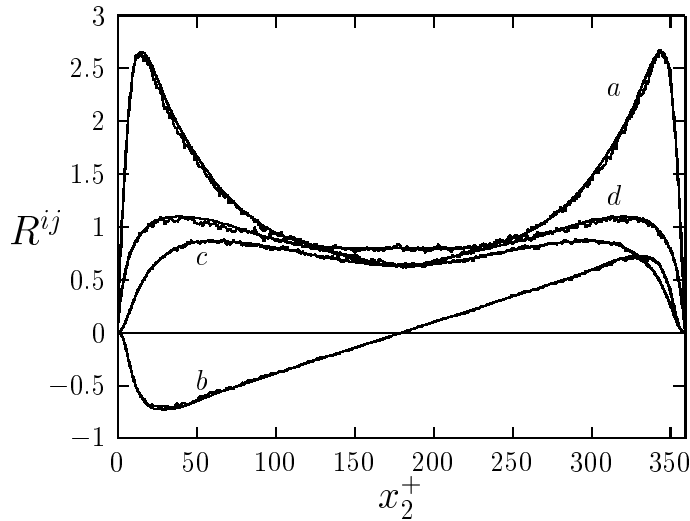,height=7.cm,angle=0.}
}
\caption{Comparison between input statistics for the Reynolds tensor $R^{ij}$, 
from DNS \cite{kim87}, and
simulated data from Montecarlo. The data are almost undistinguishable:
(a) $R^{11}$, (b) $R^{12}$, (c) $R^{22}$, (d) $R^{33}$.}
\label{fig5.fig}
\end{figure}

\begin{figure}[hbtp]\centering
\centerline{
\psfig{figure=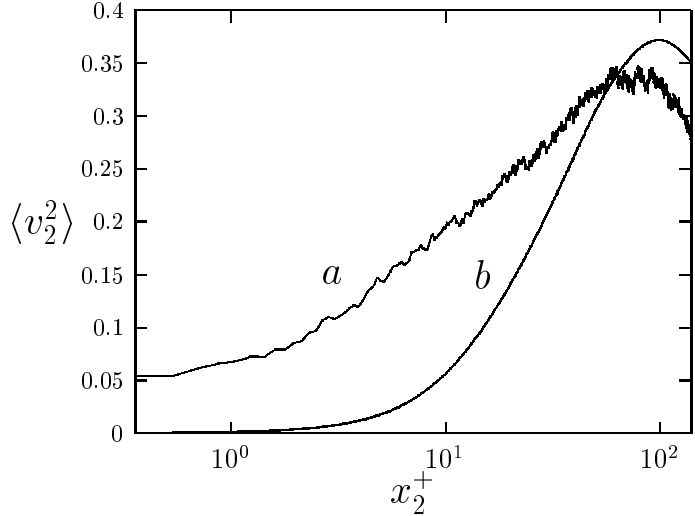,height=7.cm,angle=0.}
}
\caption{
Comparison between the
Montecarlo simulation results for $\langle v_2^2\rangle$ profile (a), and 
the homogeneous turbulence estimate for the same quantity (b).
}
\label{fig8.fig}
\end{figure}

As a validation, Figure \ref{fig5.fig} shows
that the model reproduces, in the fluid particle case, the
input statistics. Furthermore, the well-mixed
condition has been verified: despite the possible
numerical complications arising from the
presence of a multiplicative noise, a uniform passive tracer 
concentration profile is preserved in time and no tracer deposition on the walls
takes place. 
In fig. \ref{fig8.fig} we give the profile of the fluctuation amplitude
for the normal velocity of a particle with $\tau_S=60$.
The Montecarlo data strongly differ from the profile obtained from the homogeneous isotropic
turbulence estimate provided by Eqn. (\ref{corr_v0}) and illustrate the difficulty in the
a-priori determination of a reference PDF $\rho_L(\v,\x,t)$ in one-fluid Lagrangian models
(see discussion at the end of
Section VI).

\begin{figure}[hbtp]\centering
\centerline{
\psfig{figure=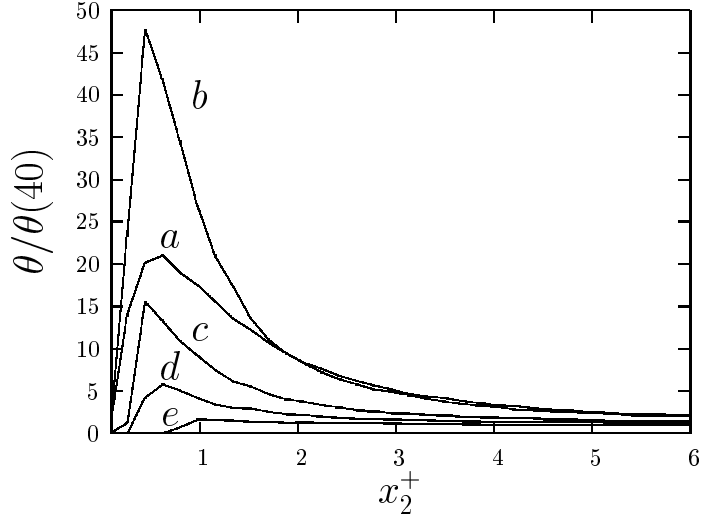,height=7.cm,angle=0.}
}
\caption{Concentration profile $\theta$ vs. $x^+_2$. 
(a) $\tau_S=3$; (b) $\tau_S=10$; 
(c) $\tau_S=20$; (d) $\tau_S=30$; (e) $\tau_S=60$.
}
\label{fig6.fig}
\end{figure}

\begin{figure}[hbtp]\centering
\centerline{
\psfig{figure=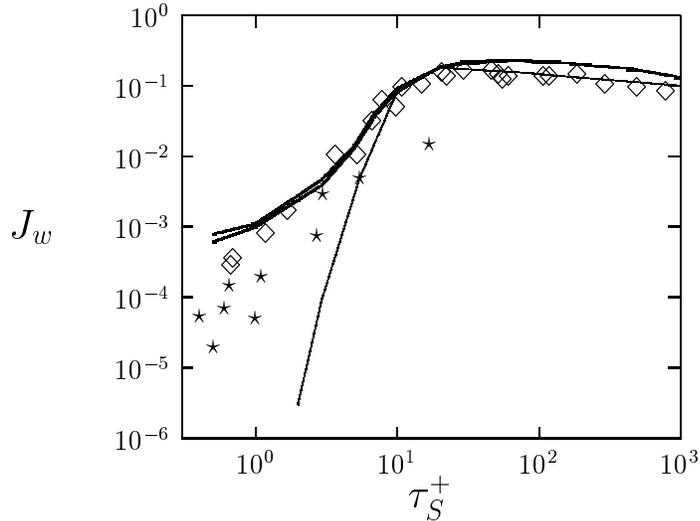,height=7.cm,angle=0.}
}
\caption{
Comparison of experimental data on particle deposition by Liu and Agarwal ($\Diamond$, Ref. 48)
and by Wells and Chamberlin ($\star$, Ref. 47), with the results of our Montecarlo simulations 
(thick lines).
The lower line corresponds to simulations without taking into account the effect of the Saffman 
lift. The thin line is the result of simulations from a one-fluid model (Ref. 39).
}
\label{fig7.fig}
\end{figure}

In Fig. \ref{fig6.fig} we give account of the particle concentration build-up in the
near wall regions.
The peak height appears to increase with $\tau_S$ 
up to $\tau_S\simeq 10$ and to decrease afterwards;
the same decrease was observed in \cite{narayanan03}. 
In agreement with both \cite{brooke94} and \cite{narayanan03}, 
and in contrast with the one-fluid model in \cite{reynolds99},
we observe that the concentration maximum occurs
in the viscous sublayer at $x_2^+\lesssim 1$.
Conversely, numerical data on the peak height present in literature show a definite scatter; 
anyway, our
data are closer to those in \cite{brooke94} than in \cite{narayanan03}, with an over-estimation
of the order of $50\%$ with respect to the first. 
\vskip 5pt
As regards particle deposition, we have studied the dependence on $\tau_S$ of the deposition
flux
\beq
J_w=\frac{L_c N_d}{T_{sim} N_{tot}}
\nonumber
\eeq
being $L_c$ the channel width, $N_d$ the number of deposited
particles in the simulation time $T_{sim}$ and $N_{tot}$ the
total number of particle simultaneously present in the channel.
In our simulations, we consider a particle deposited, when its distance from a wall is
smaller than its radius $d/2$. Assuming that air is the suspending medium, we 
fix for the density ratio the value $\gamma=1000$, and for the viscosity $\nu=0.15\cm^2/{\rm s}$;
from relation $\tau_S=1/18\,\gamma d^2/\nu$, the particle diameter will then be, in wall units:
$$
d\simeq 0.134\tau_S^\frac{1}{2}
$$
Our results are shown in Fig. \ref{fig7.fig} and compared with experimental data by
\cite{wells67,liu74}, and with an example of one-fluid Lagrangian model 
\cite{reynolds99}. 
The agreement is good with respect to the data in \cite{liu74}, apart of
a slight over-estimation in the range $\tau_S>10$. On the contrary, our model performs 
much better than the one-fluid model in the range $\tau_S<10$. As expected, the effect 
of the Saffman lift is felt only in the range $\tau_S<10$; in any case, 
the contribution to both deposition and particle accumulation appears to be small.

\vskip 20pt
\section*{XI. Conclusions}
\vskip 5pt
We have studied the statistical properties of trajectories extracted from a random velocity 
field with non-zero correlation time, analyzing the conditions for a Markovian approximation 
of the Lagrangian velocity. Our result is that a generalized form of the Thomson-87 well-mixed 
condition \cite{thomson87} can be derived also in the case of random fields, provided their
velocity structure functions scale linearly at small space-time separations. 
In the incompressible case, the Markovian approximation for the Lagrangian velocity
defines a Lagrangian model obeying the Thomson-87 well-mixed condition, with uniform 
concentration PDF given by the Eulerian one-point PDF for the random field. 

Depending on the circumstances, a random field based approach to Lagrangian modelling
may be advantageous. 
In the compressible case, knowledge of the one-point Eulerian PDF for the random field is 
sufficient to determine the coefficients of the associated Lagrangian model.
The Thomson-87 approach, instead, requires knowledge of the particular Lagrangian PDF 
(indicated in \cite{thomson87} with $g_a$) which originates from an initial concentration 
profile equal to the instantaneous local fluid density, and which does not necessarily coincide
with $\rho_E$.
Solid particle transport is an example in which implementation of 
the Thomson-87 approach is not straightforward, unless ad-hoc hypotheses are made on 
the Lagrangian statistics.

A second advantage of this approach concerns the non-uniqueness
problem: knowledge of the two-point Eulerian correlations completely fixes the form
of the Lagrangian model, which is of interest for
turbulent flows in complex geometry, where it is not clear 
which model satisfying the well-mixed condition, should be choosen.
The relation of some of the non-unique terms with helicity 
\cite{reynolds02a} and rotation \cite{sawford99}
is confirmed, and additional terms associated with strain have been identified.
Similar approaches, in which DNS informations on the
two-point Eulerian correlations are used to determine the form of Lagrangian models,
have been recently adopted in \cite{pope02}. 
Alternative formulations for the treatment of the non-uniqueness exist, in which 
the Lagrangian acceleration is modelled by a Langeving equation, on the same 
footing of the Lagrangian velocity  (second order models \cite{sawford91a}). In these 
models, however, the non-uniqueness problem is only displaced to the higher order
acceleration. 

%
%
A third advantage of the random field approach concerns situations in which it is
difficult to characterize an inertial range, and in which
concepts like the constant $C_0$ cease to be meaningful (e.g. in the buffer region of a
turbulent boundary layer).
Comparing our approach with the Thomson-87 technique, the main difference is,
to lowest order in the SO(3) expansion, the form of the noise and the parameter $a$ taking 
the place of $C_0$. 
Both noise expressions require knowledge of quantities estimated from large scale 
features of the flow: the viscous dissipation $\flux$ and the Eulerian correlation 
time $\tau_E$, whose relative dependence (as the one between $a$ and $C_0$)
is not an intrinsic characteristic of the models. 
In our approach, however, a precise relation can be obtained between 
the parameter $a$ and the ratio of the Lagrangian to the Eulerian correlation time
$\tau_L/\tau_E$, which is valid also when the Reynolds number is low.
Using for this ratio the value obtained in \cite{yeung89},
we obtain $a\simeq 0.65$.

We have tested our model, with isotropic noise and Gaussian statistics, to study 
solid particle transport both in homogenous isotropic turbulence and in 
channel flow geometry. 

In homogeneous isotropic turbulence, the correct 
renormalization for the correlation time for the fluid velocity along the
solid particle trajectories have been obtained without resorting to ad-hoc
parametrizations. The form of Eqn.  (\ref{moteq_sfv}) descends directly from
the random field and Markovianization along trajectories [see Eqns. (\ref{2.2})
and (\ref{6.5})]. This illustrates the importance of the parameter $a$ in providing
the most simple characterization of space correlations in the turbulent flow. It is 
important to stress that, had we not taken its contribution into account, 
Eqn. (\ref{moteq_sfv}) would have been unable to reproduce the
anisotropy of the time correlations. 

An interesting aspect we have observed is satisfaction of the ergodic property in 
solid particle transport by
homogeneous turbulence, this, despite compressibility of the solid particle flow. 
It is not clear whether this is an artifact of the model; in 
any case, it is a non-trivial effect since the ergodic property can be shown to be
violated by very simple compressible flows [see Eqns. (\ref{7.3.1}-\ref{7.3.3}) and
discussion therein].

In channel flow geometry, we have found good agreement with experimental
data on particle deposition \cite{wells67,liu74}, and partial agreement with 
numerical data on near wall accumulation \cite{brooke94,narayanan03}.
We stress that these results have been obtained without any parameter fitting, apart 
of the choice $a=0.65$ inferred from \cite{yeung89}.

Clearly, a Gaussian model with isotropic noise cannot account for the effect of
coherent structures and intermittency, which are an important feature
of turbulence in channel flows. Imposing the appropriate form for the one-point
PDF and going to higher orders in the SO(3) expansion allows consideration
of these effect. 
Preliminary analysis suggests that inclusion in the model of non-Gaussianity,
noise anisotropy and non-unique terms, strongly affects particle deposition 
and transport in 
wall turbulence. Modelling the structure of the turbulent correlations, based
on empirical considerations, appears to lead to models that perform worse,
compared to the data, than the simple isotropic Gaussian model, a situation 
similar to that observed in \cite{flesch92}.
This suggests that careful consideration of the structure of the 
turbulent correlation, based on DNS data, may be necessary; this will
be part of a different publication.

Some issues remain to be clarified as regards the definition of a 
random field purely in terms of its local properties. The global extension
provided by Eqn. (\ref{2.9}), in which the random field is assumed ''Markovian'' 
along rectilinear cuts in space-time is only one of the possibilities. 
This choice produces effects on the form of the trajectories, which
can be accounted for only in the non-Markovian approach described in 
Section VIII. (Markovianization along trajectories corresponds 
to considering only local properties of the random field).  An open question 
remains which global structure of a random field would lead, for fixed
local structure, to transport properties which are approximated best
by a Markovian Lagrangian model. This, beside 
understanding whether the assumption in Eqn. (\ref{2.9}), which leads
in the Gaussian case
to exponential scaling of the random field correlations, fits
the  turbulent structure in appropriate way.
For the choice provided by 
Eqn. (\ref{2.9}), and for homogeneous isotropic conditions and Gaussian statistics, 
the error in the ratio $\tau_L/\tau_E$ corresponding to $a\simeq 0.65$
appears to be of the order of $15\%$ in defect.

Related to this issue is the fact that 
consideration of long-lived coherent structures, corresponding 
to small values of $\tau_L/\tau_E$ and 
to large errors in the Markovian approximation, is probably out
of the range of applicability of our model.
An alternative strategy, which would allow taking into account long-lived coherent
structures, is the non-Markovian approach described in Section VIII. The noise and
drift terms, however, would have to be rederived including the condition at the emission point
following Eqns. (\ref{8.14}) and (\ref{8.17}).
\vskip 5pt
Extension of the present approach beyond one-point statistics is possible in principle, 
but is limited by the unphysical scaling of the random field structure function at small
separations. 
Only concentration fluctuations on the scale of
a correlation length could then be taken into account.

\vskip 5pt

\vskip 15pt
\noindent{\bf Aknowledgemnt}: We aknowledge support by Agenzia2000 grant
CNR C002509\_003.
\vskip 20pt

\appendix

\vskip 20pt
\centerline{ \bf Appendix A: non-Gaussian case}

\vskip 5pt
A symmetric one-dimensional distribution with unitary variance and
kurtosis larger than three can be modelled by means of a bi-Gaussian:
$$
P(x)=\frac{\alpha}{\pi^\frac{1}{2}}\exp(-x^2)+\frac{(1-\alpha)}{(2\pi\beta)^\frac{1}{2}}
\exp(-\frac{x^2}{2\beta})
\eqno(A1)
$$
where
$$
\alpha=\frac{4\langle x^4\rangle-12}{4\langle x^4\rangle-9}
\quad
{\rm and}
\quad
\beta=\frac{2}{3}\langle x^4\rangle-1;
\eqno(A2)
$$
parameterize the strength of the kurtosis $\langle x^4\rangle$.
From here, we can obtain the expression for an isotropic non-Gaussian velocity distribution:
$$
\rho_E=\frac{1}{(\pi u_T^2)^\frac{3}{2}}\Big[\alpha\exp(-\frac{u^2}{u_T^2})
+\frac{1-\alpha}{(2\beta)^\frac{3}{2}}\exp(-\frac{u^2}{2\beta u_T^2})\Big]
\eqno(A3)
$$
and for an anisotropic distribution, in which one of the velocity components, in the diagonal
reference frame for the Reynolds tensor, is non-Gaussian:
$$
\rho_E=\alpha\rho_1+(1-\alpha)\rho_2=
\rho_x[\alpha\rho_{y\smalun}+(1-\alpha)\rho_{y\smaldu}]\rho_{z}
\eqno(A4)
$$
where:
$$
\rho_x=\frac{1}{(2\pi\hat R^{11})^\frac{1}{2}}\exp(-\frac{\hat u_1^2}{2\hat R^{11}}),
\qquad
\rho_{z}=\frac{1}{(2\pi\hat R^{33})^\frac{1}{2}}\exp(-\frac{\hat u_1^2}{2\hat R^{33}}),
$$
$$
\rho_{yi}=\frac{1}{(2\pi\hat R_\smali^{22})^\frac{1}{2}}
\exp(-\frac{\hat u_2^2}{2\hat R_\smali^{22}})
\quad i=1,2
\eqno(A5)
$$
with $\hat R_\smalun^{22}=\hat R^{22}/2$, $\hat R_\smaldu^{22}=\beta\hat R^{22}$, and
the hat indicating the diagonal reference frame.

Let us calculate explicitly the
drift terms in the case of Eqns. (A4-A5). Substituting into Eqn. (\ref{3.4}), 
we find immediately
that $\bar A$ is given by the superposition of the contributions from each of the Gaussians
$\rho_1$ and $\rho_2$:
$$
\bar A_\mu^{\ i}\d x^\mu=-\frac{1}{2\rho}\langle\d w^i\d w^j\rangle[
\alpha\rho_\smalun S^\smalun_{jk}+(1-\alpha)\rho_\smaldu S^\smaldu_{jk}]
u^k
\eqno(A6)
$$
(notice the absence of the hats; it is not necessary here to work in the diagonal reference
frame).
The contribution from the $\Phi$ and $\Psi$ terms has a more complicated form. Let us take
the laboratory frame with the inhomogeneity direction along $x^2$ (the usual channel flow
geometry in which $x^1$ is the mean flow direction). We use the ansatz:
$$
\Phi_\mu^{\ i}=\alpha\Phi_{\mu\smalun}^{\ i}
+(1-\alpha)\Phi_{\mu\smaldu}^{\ i}+\Delta\Phi_\mu^{\ i},
\qquad
\Delta\hat\Phi_\mu^{\ i}=\hat F_\mu\delta^i_2
\eqno(A7)
$$
where $\Phi_\smalun$ and
$\Phi_\smaldu$ give the form of $\Phi$ in the case $\rho_E=\rho_\smalun$ and
$\rho_E=\rho_\smaldu$.
Substituting into Eqn. (\ref{3.5}) and using Eqns. (A4-A5), we obtain:
$$
\partial_{\hat u^2}\hat F_\mu=-\delta^2_\mu\hat\partial_2\alpha(\rho_1-\rho_2)
\eqno(A8)
$$
leading to the result in the laboratory reference frame:
$$
\Delta\Phi_\mu^{\ i}=-\delta^2_\mu \Omega^i_{\ 2}
\frac{\rho_x\rho_y}{(2\pi)^\frac{1}{2}}\partial_2\alpha
\int^{\hat u_2/\sqrt{R_\smalun^{22}}}
_{\hat u_2/\sqrt{R_\smaldu^{22}}}\d\hat u\ \ex^{-\hat u^2/2}
\eqno(A9)
$$
where $\Omega^i_{\ 2}$ is the rotation matrix defined through $u^i=\Omega^i_{\
j}\hat u^j$, i.e.
$\Omega^i_{\ j}=\e^i\cdot\hat\e_j$.

Analogous procedure is followed to obtain the $\Psi$ term. From Eqn. (\ref{3.6}), in 
analogy with Eqn. (\ref{3.11}), write:
$$
\psi^i=\alpha\psi^i+(1-\alpha)\psi^i+\Delta\psi^i
\qquad
\Delta\hat\psi^i=\delta^i_2G
\eqno(A10)
$$
where $2\partial_{u^i}\psi^i=-\Delta\Phi_i^i=-\hat F_2$, and decompose $\Psi_j^{\ i}$ in
analogous way. From Eqn. (A9), we obtain immediately:
$$
G=\frac{\rho_x\rho_y}{2(2\pi)^\frac{1}{2}}\Omega^2_{\ 2}
\partial_2\alpha\int_{-\infty}^{\hat u_2}\d\hat u
\int^{\hat u/\sqrt{R_\smalun^{22}}}
_{\hat u/\sqrt{R_\smaldu^{22}}}\d x\ \ex^{-x^2/2}
\eqno(A11)
$$
and, substituting again into Eqn. (\ref{3.11}) $[$in real space: 
$\Psi_j^{\ i}=\delta^i_j\partial_{u^k}\psi^k-\partial_{u^j}\psi^i]$, we find:
$$
\Delta\Psi_j^{\ i}=\delta^i_j\partial_{\hat u^2}G-\Omega^i_{\ 2}\Omega_j^{\ l}\partial_{\hat u^l}G
\eqno(A12)
$$
Assuming that the statistics of $\d\w$ be independent of the velocity, as we have done in 
section II, has the consequence that all of the non-Gaussianity is contained in the
drift. The small scale structure of the correlations, associated with $\d\w$ remain 
therefore Gaussian. This breaks, for large values of the kurtosis, the direct relationship
between the drift coefficients and the correlation lengths \cite{maurizi00}. It is easy to see what happens
looking at Eqns. (A4) and (A6). Whenever the value of $u$ goes above $u_T$, 
the slowly decaying $\rho_\smaldu$ becomes dominant in Eqn. (A6) and leads to a reduced
decay rate for the fluctuation that can thus slowly grow to produce a burst. We thus have 
a hierarchy of time-scales (focus for simplicity on variations along time):
$$
\tau_E\longrightarrow
{\rm\ Time\ scale\ for\ background\ (}u\sim u_T)
$$
$$
\beta\tau_E\longrightarrow
{\rm\ Time\ scale\ for\ bursts\ (}u\sim\beta^\frac{1}{2}u_T)
$$
$$
\beta^2\tau_E\longrightarrow
{\rm\ Spacing\ between\ bursts}
$$
This difference between the time-scale for bursts and background fluctuations is not physically
meaningful in general. This has the effect of overshooting the burst contribution to
turbulent dispersion, which is estimated by the product of the probability $\beta^{-1}$ of 
a burst, its time scale and the square of its velocity scale: 
$$
\beta^{-1}\times\beta\tau_E\times\beta u_T^2=\beta u_T^2\tau_E.
\eqno(A13)
$$
(In comparison, the background contribution is $u_T^2\tau_E$, and, if the time-scales for burst
and  background were the same, the background and burst contributions would be identical).

\vskip 20pt
\centerline{ \bf Appendix B: Spherical tensors and the SO(3) technique}
\vskip 5pt
A symmetric two-index tensor function can be decomposed in spherical tensors in the form
$$
\delta^{ij}Y_J(\x),
\qquad
\partial^i\partial^jY_J(\x),
\qquad
x^ix^jY_J(\x),
\qquad
(x^i\partial^j+x^j\partial^i)Y_J(\x),
$$
$$
x_n(\epsilon^{jnm}x^i+\epsilon^{inm}x^j)\partial_mY_J(\x)
\quad
{\rm and}
\quad
x_n(\epsilon^{jnm}\partial^i+\epsilon^{inm}\partial^j)\partial_mY_J(\x)
\eqno(B1)
$$
where $Y_J(\x)$ is a $J$-order polynomial in the components $x_i$:
$$
Y_J(\x)=y^{i_1i_2...i_J}x_{i_1}x_{i_2}...x_{i_J}
\eqno(B2)
$$
and $y^{i_1i_2...i_J}$ is traceless with respect to any pair of indices \cite{arad99}.
In consequence of this, the spherical tensors in Eqn. (B1) will be polynomials of order
$L=J$, $J-2$, $J+2$, $J$, $J+1$ and $J-1$ respectively. In the case of the noise tensor
$\langle\Delta w^i\Delta w^j\rangle$, we have the additional symmetry with respect to
spatial inversion, which imposes the condition that $L$ be even. This
implies $J$ even for the first four spherical tensors and $J$ odd for the last two.
Limiting the analysis to $J\le 2$, we notice immediately that the last spherical 
tensor in Eqn. (B1) disappears. Similarly the $J=1$ contribution from 
$x_n(\epsilon^{jnm}x^i+\epsilon^{inm}x^j)\partial_mY_J(\x)$
is absent due to incompressibility; writing $Y_1(\x)=y_mx^m$:
$$
\partial_ix_n(\epsilon^{jnm}x^i+\epsilon^{inm}x^j)\partial_mY_1(\x)=5\epsilon^{jnm}x_ny_m=0
$$
which imposes $y_m=0$. We are thus left only with the $J=0$ and $J=2$ contributions.

From Eqn. (\ref{4.3}), the $J=0$ and $J=2$ contributions to $B^{ij}$ will have the form:
$$
u_TB_0^{ij}(\x)=a|\x|\delta^{ij}+\hat a\frac{x^ix^j}{|\x|}]
\eqno(B3)
$$
$$
u_tB_2^{ij}(\x)=\frac{4b^{lm}}{|\x|}x_lx_m\delta^{ij}
+2c^{lm}|\x|\partial_i\partial_jx_lx_m
+\frac{d^{lm}}{|\x|^3}x_lx_mx^ix^j
$$
$$
+\frac{e^{lm}}{2|\x|}(x^j\partial^i+x^i\partial^j)x_lx_m
\eqno(B4)
$$
Applying the incompressibility condition $\partial_iB^{ij}=0$ leads to the equations:
$$
\begin{cases}
\hat a=-\frac{a}{3}
\\
8b^{lm}+4c^{lm}+8e^{lm}=0
\\
3d^{lm}-4b^{lm}-e^{lm}=0
\end{cases}
\eqno(B5)
$$
Substituting the solution to Eqn. (B5) into Eqns. (B3-B4) leads to Eqn. (\ref{4.4}).
\vskip 5pt
We give next the expressions for the spherical tensors contributing to an antisymmetric
two-index tensor:
$$
\epsilon^{ijk}x_kY_J
\qquad
\epsilon^{ijk}\partial_kY_J
\quad
{\rm and}
\quad
(x^i\partial^j-x^j\partial^i)Y_J
\eqno(B6)
$$
In the case of the antisymmetric part of the correlation $C_A$, we have the
additional property of antisymmetry with respect to spatial inversion, which implies
that $J$ be even for the first two and odd for the last.
\vskip 5pt
In the case of a vector field, an analogous decomposition can be obtained in terms of 
spherical vectors in the form:
$$
x^iY_J(\x),
\qquad
\partial^iY_J(\x)
\quad
{\rm and}
\quad
\epsilon^{ijk}x_j\partial_kY_J(\x)
\eqno(B7)
$$
If the vector field does not have an axial component, only the first two spherical vectors 
can contribute. If we have axial symmetry, identified by a direction $\u$, the 
tensors $y^{i_1i_2...i_l}$  entering Eqn. (B2) will be zero trace symmetrized products 
of components $u^i$ and of the identity matrix $\delta^{ij}$. The first spherical vectors
$x^iY_J(\x)$ and $\partial^iY_J(\x)$ are respectively:
$$
\x,
\qquad
(\u\cdot\x)\x,
\qquad
((\u\cdot\x)^2-\frac{1}{3}|\u|^2|\x|^2)\x
\eqno(B8)
$$
and
$$
0,
\qquad
\u,
\qquad
(\u\cdot\x)\u-\frac{1}{3}|\u|^2\x
\eqno(B9)
$$

%

\vskip 20pt
\centerline{ \bf Appendix C: Conditional random field statistics}
\vskip 5pt
We want to calculate conditional velocity moments in the form
$$
\langle u^i(\x_0,t_0)u^j(\x_0,t_0)...|\u(\x_1,t_1),\u(\x_2,t_2),...\rangle
\eqno(C1)
$$
Let us indicate, for $l=0,1,...n$, $\U_l=\u(\x_l,t_l)$, and 
${\bf C}_{lm}=\langle\U_l\U_{m}\rangle$, assuming for simplicity a symmetric correlation tensor.
For a Gaussian random field, the velocity correlations at points $\{t_l,\x_l\}$ are obtained
from the generating function
$$
\tilde\rho(\{{\boldsymbol{\eta}}_l\})=
\exp\Big(-\frac{1}{2}\sum_{l,m=0}^n
{\boldsymbol{\eta}}_l
\cdot{\bf C}_{lm}\cdot
{\boldsymbol{\eta}}_{m}\Big)
\eqno(C2)
$$
Let us introduce the marginal PDF
$$
\rho(\{\U_l,i=1,...n\})=
{\cal N}\exp\Big(-\frac{1}{2}\sum_{l,m=1}^n\U_l\cdot{\bf D}_{lm}\cdot\U_m\Big)
\eqno(C3)
$$
where ${\cal N}$ is the normalization and
${\bf D}_{lm}$ is the inverse of the restriction of ${\bf C}_{lm}$ to $l,m=1,...n$.
We shall indicate in the following this restriction with a prime:
$$
{\sum_{lm}}'\equiv \sum_{l,m=1}^n,
\qquad
\{\U'_l\}\equiv\{\U_l,l=1,...n\},
\quad
{\rm etc.}
\eqno(C4)
$$
The generating function for $\U_0$ conditioned to $\U_l$, $l=1,...n$ is obtained by inverse 
Fourier transforming $\tilde\rho(\{{\boldsymbol{\eta}}_l\})$ with respect to 
$\{{\boldsymbol{\eta}}'_l\}$ in $\{\U'_l\}$ and dividing by the marginal PDF $\rho(\{\U'_l\})$:
$$
\tilde\rho({\boldsymbol{\eta}}_0|\{\U'_l\})=\frac{1}{\rho(\{\U'_l\})}
\int{\prod_l}'\d^3\eta_l\exp\Big(-{\rm i}{\sum_l}' {\boldsymbol{\eta}}_l \cdot \U_l
$$
$$
-\frac{1}{2}{\sum_{lm}}'
{\boldsymbol{\eta}}_l\cdot{\bf C}_{lm}\cdot{\boldsymbol{\eta}}_{m}
-{\sum_l}'{\boldsymbol{\eta}}_0\cdot{\bf C}_{0l}\cdot{\boldsymbol{\eta}}_l
-\frac{1}{2}{\boldsymbol{\eta}}_0\cdot{\bf C}_{00}\cdot{\boldsymbol{\eta}}_0 \Big)
\eqno(C5)
$$
From here we can calculate the conditional moments in Eqn. (C1).
We calculate first the mean velocity in $\{t_0,\x_0\}$ given velocities $\U_l$ in $\{t_l,\x_l\}$
$l=1,...n$:
$$
\langle\U_0|\{\U'_l\}\rangle=\frac{1}{\tilde\rho(0|\{\U'_l\})}
\frac{\partial\tilde\rho({\boldsymbol{\eta}}_0|\{\U'_l\})}
{\partial{{\rm i}{\boldsymbol{\eta}}_0}}\Big|_{{\boldsymbol{\eta}}_0=0}
$$
$$
=\frac{{\sum_r}'{\bf C}_{0r}}
{\rho(\{\U'_l\})\tilde\rho(0|\{\U'_l\})}
\cdot \int{\prod_l}'\d^3\eta_l\, {\boldsymbol{\eta}}_r \exp\Big(-\frac{1}{2}
{\sum_{lm}}'
{\boldsymbol{\eta}}_l\cdot{\bf C}_{lm}\cdot{\boldsymbol{\eta}}_{m}
-{\rm i}{\sum_l}'
{\boldsymbol{\eta}}_l\cdot\U_l\Big) 
\eqno(C6)
$$
Carrying out the Gaussian integrals, we obtain the result:
$$
\langle\U_0|\{\U'_l\}\rangle={\sum_{lm}}'{\bf C}_{0l}\cdot{\bf D}_{lm}
\cdot\U_{m}
\eqno(C7)
$$
The calculation of the second conditional moment is analogous and the result is:
$$
\langle\U_0\U_0|\{\U'_l\}\rangle=-\frac{1}{\tilde\rho({\boldsymbol{\eta}}_0|\{\U'_l\})}
\frac{\partial^2\tilde\rho({\boldsymbol{\eta}}_0|\{\U'_l\})}
{\partial{\boldsymbol{\eta}}_0
\partial{\boldsymbol{\eta}}_0}\Big|_{{\boldsymbol{\eta}}_0=0}
$$
$$
={\bf C}_{00}-{\sum_{lm}}'{\bf C}_{0l}\cdot {\bf D}_{lm}\cdot{\bf C}_{m0}
+\langle\U_0|\{\U'_l\}\rangle\langle\U_0|\{\U'_l\}\rangle
\eqno(C8)
$$

\vskip 20pt

\end{document}